\documentclass[useAMS,usenatbib]{mn2e}
\usepackage{epsfig,appendix,array,rotating,pdflscape,array,longtable,colortbl}
\usepackage{multirow}
\usepackage{booktabs}
\usepackage{lscape, amsmath} 

\usepackage{relsize}
\usepackage{fixltx2e}
\usepackage{amssymb}

\newcommand{\hi}{H\,{\sc i}}

\newcommand{\prim}{$^{\prime}$}
\newcommand{\prin}{$^{\prime\prime}$}
\newcommand{\aprox}{${\sim}$}

\newcommand{\km}{km\,s$^{-1}$}
\newcommand{\degree}{$^{\circ}$}
\newcommand{\halpha}{H${\alpha}$}
\newcommand{\cg}{CGCG\,097--}
\newcommand{\msolar}{M$_{\odot}$}

\newcommand{\LB}{$L_{B}$}

\newcommand{\hdos}{H$_{2}$}
\newcommand{\defhi}{H\,{\sc i} deficiency}

\newcommand{\msolaryr}{M$_{\odot}$\,yr$^{-1}$}
\newcommand{\ico}{$I$(CO)}
\newcommand{\hto}{CO ($J=2 \rightarrow 1$)}
\newcommand{\hoz}{CO ($J=1 \rightarrow 0$)}
\newcommand {\apgt} {\ {\raise-.5ex\hbox{$\buildrel>\over\sim$}}\ }
\newcommand {\aplt} {\ {\raise-.5ex\hbox{$\buildrel<\over\sim$}}\ }
\newcommand{\dhi}{$\mathit{Def}_{HI}$}
\newcommand{\hd}{$\mathit{Def}_{H_2}$}
\newcommand{\me}{$\mathit{(M_{H_2})_e}$}
\newcommand{\mo}{$\mathit{(M_{H_2})_o}$}
\setlongtables

\title[]{ CO in late--type galaxies within the central region of Abell\,1367}
\author[]{T. C. Scott$^{1,2}$\thanks{E-mail:
t.c.scott@herts.ac.uk (TS)}, A. Usero$^{3}$,  E. Brinks$^{2}$, A. Boselli$^{4}$, 
L. Cortese$^{5}$, and H. Bravo--Alfaro$^{6}$
\\
$^{1}$Instituto de Astrof\1sica de Andaluc\1a (CSIC), Apartado 3004, 18080 Granada, Spain\\
$^{2}$Centre for Astrophysics Research, University of Hertfordshire, College Lane, Hatfield, AL10 9AB, UK\\ 
$^{3}$Observatorio Astron\'omico Nacional, C/Alfonso XII 3, 28014 Madrid, Spain\\
$^{4}$Laboratoire d'Astrophysique de Marseille, OAMP, Universit$\acute{e}$ Aix--Marseille 8 CNRS, 38 Rue Fr$\acute{e}$d$\acute{e}$ric Joliot--Curie, 13388 Marseille, France\\
$^{5}$ESO, Karl--Schwarzschild--Str.\ 2, 85748 Garching bei M\"unchen, Germany\\
$^{6}$Departamento de Astronom\'\i a, Universidad de Guanajuato, Apdo.\ Postal 144, Guanajuato 36000, Mexico}

\begin{document}
\date{Accepted. Received ; in original form }


\maketitle

\label{firstpage}

\begin{abstract}
We present $^{12}$\hoz\ and $^{12}$\hto\ spectra for 19 bright, late--type galaxies (spirals) in the central region of the galaxy cluster Abell 1367 (\textit{z} = 0.02) from observations made with the IRAM 30--m telescope. All 19 spirals were observed at the position of their optical center and for a subset, at multiple positions. For each spiral the integrated \hoz\ intensity from the central pointing, in few cases supplemented with intensities from offset pointings, was used to estimate its molecular hydrogen mass and \hdos\ deficiency. Accepting the considerable uncertainties involved in determining \hdos\ deficiencies, spirals previously identified by us to have redder colours and higher \hi\ deficiencies as a result of environmental influence, were found to be more \hdos\ deficient compared to members of the sample in less advanced evolutionary states. For eight of the observed spirals multiple pointing observations were made to investigate the distribution of their molecular gas. For these spirals we fitted Gaussians to the CO intensities projected in a line across the galaxy. In two cases, \cg079 and  \cg102(N), the offset between the CO and optical intensity maxima was significantly larger than the pointing uncertainty and the FWHMs of the fits were significantly greater than those of the other spirals, irrespective of optical size. Both signatures are indicators of an abnormal molecular gas distribution. In the case of \cg079, which is considered an archetype for ram pressure stripping, our observations indicate the CO intensity maximum lies $\sim$ 15.6 $\pm$ 8.5 arcsec (6 kpc) NW of the optical centre at the same projected position as the \hi\ intensity maximum.
\end{abstract}


\section{INTRODUCTION}
Two principal classes of mechanisms are invoked to explain the rapid evolution of the Interstellar Medium (ISM) of late--type galaxies (spirals) in galaxy clusters: hydrodynamic interactions \citep{gun72,nuls82,bekki02,fuji04} and fast tidal encounters \citep[][]{moore96,moore99,bekki99,nata02}.  \textcolor{black}{These processes are extensively discussed and compared in \citet{bosel06a}} The sensitivity and spatial resolution of current instruments limits detailed observational studies of the effect of these types of interaction on the ISM to nearby clusters ($z$ \aplt 0.05).

Typically the radius of an unperturbed spiral's \hi\ disk extends $\sim$ 1.8 times further than the optical disk \citep{caya94,broeils97}. In contrast the molecular gas is usually concentrated in spiral arms well inside the optical disk \citep{dick92,leroy08,leroy09}, being more tightly bound gravitationally than the \hi. In the absence of a recent major interaction, the stellar, \hi, and molecular disks are expected to be, to first order, distributed axisymmetrically with the centres of the \hi\ and molecular disks coinciding with the optical centre.

In the cluster environment \hi\ deficiencies have been interpreted as evidence of past gas loss \citep{vgork04}, due to removal of \hi\ beyond a position where it is not expected to fall back into the spiral's potential and/or conversion of \hi\ to other hydrogen phases. In general to determine whether a spiral contains a normal amount of cold ISM its \hi\ and \hdos\ masses are compared to the mean \hi\ and \hdos\ content of isolated spirals of similar size and Hubble type. A spiral's deficiency (in \hi\ or
\hdos) is defined as the log of the ratio of the expected to observed gas mass \citep{hayn84,bosel97,bosel02}. The \hdos\ deficiency (\hd) of a cluster spiral is the net result of several competing processes which determine the mass of its observed \hdos\ reservoir, e.g depletion of \hdos\ by star formation, dissociation by the UV radiation emitted by high mass stars, direct removal during interactions and replenishment of \hdos\ by conversion of \hi\ to \hdos\ \citep{kurmh08,krumh09}.

We can consider star formation in spirals as being the end result of a continuous process of converting \hi\  $\rightarrow$ \hdos\  $\rightarrow$ stars. How the relative proportions of these components is altered during and as a result of either tidal or hydrodynamic interactions is poorly understood, primarily because of the limited availability and uncertainties surrounding the \hdos\ data. Models predict increased star formation rates (SFR) in cluster spirals during both hydrodynamic and tidal interactions. These models indicate both types of interaction are capable, in specific circumstances, of enhancing star formation by up to an order of magnitude \citep{mart08,kronb08,kapf08,kapf09}. Alternative explanations for \hdos\ deficiencies, include starvation under which a spiral's \hi\ halo is stripped by interaction with the cluster's intracluster medium (ICM). In this scenario, a spiral's cold molecular reservoir is depleted by a combination of normal star formation and reduced availability of \hi\ to convert to \hdos\ \textcolor{black}{\citep[e.g.][]{larson80,bekki02}.} It should be noted, however, that the time scale for starvation effects is several Gyr  \citep{bosel06a,bosel06b}.

Previous studies of molecular deficiencies in clusters, e.g. \cite{young89} for the Virgo cluster and \cite{casoli91} for the Coma super cluster, concluded that cluster spirals are not deficient in \hdos\ compared to those in the field.	But	these	and	other	 $L_{FIR}$  	selected	samples	are	biased against finding \hdos\ deficient galaxies, because of the strong correlation between $M_{H_2}$  per unit mass and $L_{FIR}$ \citep{bosel97}. Using an optically selected sample, \cite{bosel02} reached the same conclusion. These results have been used to support the argument that the observed \hi\ deficiencies in clusters primarily arise from hydrodynamic stripping on the basis that ram pressure strips \hi\ while leaving the \hdos\ component unperturbed \citep{bosel06a}. Given that star formation in cluster spirals is lower than in the field \citep{kenni83,gava02,koop04} the lack of overall \hdos\ deficiency is surprising as it implies a typical cluster spiral has suppressed star formation despite having normal \hdos\ content. To date only a few \hdos\ deficient cluster spirals have been reported  \citep[e.g.][]{fuma09}, but this may reflect the bias in most previous surveys against finding such deficiencies \citep{bosel97,bosel02}.

How ongoing interactions (tidal or hydrodynamic) impact the amount and distribution of cold molecular gas in cluster spirals is much less well studied than for the neutral atomic gas. CO images of spirals involved in strong tidal interactions range from bar--like to tri--lobal distributions \citep[e.g.][]{kauf02,iono05,cortes06,combes09}. \cite{iono05} reported off- sets between \hi\ and optical intensity maxima in 50 \% of a sample of tidally interacting galaxies but their study (their Table 9) indicates the impact on the distribution of molecular and \hi\begin{scriptsize}
•
\end{scriptsize}\ components may be quite different depending on the parameters of the interaction.

Most hydrodynamic interaction simulations do specifically model the cold molecular ISM \citep{roed05,roed07,kronb08,kapf09}. In the few cases where spatially resolved molecular observations have been utilised in combination with data from other wavelengths and sticky particle simulations they have proved extremely fruitful in distinguishing between interaction mechanisms \citep{voll01b,voll08}. It is generally thought that molecular disks remain largely undisturbed by hydrodynamic interactions \citep{lucero05, voll08}. This may, in part, be because the best studied cases are experiencing relatively low ram pressures, e.g., NGC 4522 which is $\sim$ 1 Mpc from its cluster's core, with 75 \% of the molecular gas remaining co-spatial with the optical disk and showing a SFR of 0.1 \msolaryr\  \citep{voll08}. However ESO 137-001, projected 180 kpc from the centre of Abell\,3627 appears to be experiencing an order of magnitude greater ram pressure than NGC\,4522. ESO 137-001 has a recently discovered warm molecular tail implying the interaction has resulted in a significant mass of cold molecular gas downstream of the optical disk  \citep[][]{sivanandam10,sun10}. The case of ESO 137-001 highlights that the relation between the magnitude of a hydrodynamic interaction and its effect on the distribution of disk and extra--planar molecular gas remains poorly constrained.

We are investigating the impact of the cluster environment on the ISM  of a sample of 26 optically selected bright late--type galaxies within a central radius (1.5 Mpc) of the nearby cluster A\,1367,  begun with a programme of VLA \hi\ observations (Scott et al. 2010 -- Paper I)\nocite{scott10}. The majority of  spirals from this sample imaged so far with the VLA show asymmetric \hi\ distributions and \hi\ deficiencies, implying they are  suffering significant interactions of some kind. 

In this paper we report on the  \hoz\ and \hto\ emission from 19 of the 26 galaxies from Paper I. The 7 galaxies from the Paper I sample which were not observed in CO were principally those with the smallest diameter optical disks. The most extensive previous CO observations covering the central region of A\,1367 were carried out with the NRAO Kitt Peak 12--m telescope \citep{bosel97}. Several spirals within the central region have previously been observed with the IRAM\footnote{IRAM is supported by CNRS/INSU (France), the MPG (Ger- many) and the IGN (Spain).} 30--m  \citep[e.g.][]{casoli91,dick92,bosel94,lave99}. In comparison with previous CO studies within the central region our current observations include a larger number of galaxies, often having higher angular resolution, and in cases of re-observations, almost all are more sensitive.

In Paper I we distinguished four broad evolutionary states (A--D) for the spirals using ranges of \hi\ deficiency and SDSS colour.  \textcolor{black}{We continue to use the same definitions of evolutionary state in the current paper, with the parameters given in Table \ref{estates}. The classification assumes that interaction induced removal/consumption of ISM gas (\hi) will eventually lead to suppression of star formation and a change of optical colour.}

\begin{table}
\centering
\begin{minipage}{140mm}
\caption{\textcolor{black}{Evolutionary states of spirals}}
\label{estates}
\begin{tabular}{@{}cll@{}}
\hline
State &\hi\ &Colour  \\
&deficiency &SDSS \textit{g--i}   \\
\hline 
A& $<$0.7&$<$0.76 \\ 
B& $<$0.7& $ \geq $0.76 and $<$1.1 \\ 
C & $<$0.7& $\geq$1.1 \\ 
D & $\geq$0.7& any \\ 
\hline
\end{tabular}
\end{minipage}
\end{table}

Based on a redshift to A\,1367  of 0.022 and assuming $\Omega_M =0.3$, $\Omega_\Lambda =0.7$ \ and \ $H_0$ =72\,\km\,Mpc$^{-1}$ \citep{sperg07} the distance to the cluster is 92 Mpc with an  angular scale of 1 arcmin $\approx$ 24.8 kpc. All $\alpha$ and $\delta$ positions referred to throughout the paper are J2000.0.

In section \ref{secobs} we describe the observations and reduction. The observational results are set out in section \ref{results} and discussed in section \ref{discuss}, followed by concluding remarks in section \ref{conclusion}.

\section{OBSERVATIONS}
\label{secobs}

The observations were carried out in July 2008 using the IRAM 30--m telescope at Pico Veleta, Spain, with simultaneous observations of $^{12}$\hoz; rest frequency 115.271 GHz; 22 arcsec beam and $^{12}$\hto; rest frequency 230.538 GHz; 11 arcsec beam, with the receivers tuned to the redshifted frequencies corresponding to the optical velocity of each source. The main observational parameters are set out in Table  \ref{obs}. We checked calibration against the CN(1-0) line in IRC+10216. The calibration was consistent within 15-20\% with the intensities measured by \cite{mauersberger89}. After every two scans ($\sim$ 15 minutes) a chopper wheel calibration was carried out. Pointing was checked at between 1 and 2 hour intervals using the broadband continuum from Mars, 3C273, 4C29.45 and 4C01.28. The IRAM 30--m webpage indicates an average pointing accuracy of $\sim$ 2 arcsec or better. This value is in good agreement with the differences in pointing corrections that we found between consecutive pointing measurements. All observations were made in wobbler switching mode with a frequency of 0.5 Hz and \textcolor{black}{an on--off  throw of $\pm$220 arcsec in azimuth.} For the galaxy with the largest optical major axis, \cg129(W), the ON--OFF throw is estimated to leave a separation between the outer edge of the optical disk and outer edge of the 2.6 mm beam in the off position of $\sim$ 55kpc. The 1 MHz and 4 MHz backends were used at 2.6 mm and 1.3 mm respectively. This gave a velocity resolution for the raw data of $\sim$ 2.6 \km\ at 2.6 mm and $\sim$ 5.2 \km\ at 1.3 mm, but for our analysis the channels were binned to a velocity width of	10 or 20 \km. The	weather	conditions	were	generally fairly good with zenith opacity $\tau$  ranging from 0.1 to 0.4 at 225 GHz.

\begin{table*} 
\centering
\begin{minipage}{140mm}
\caption{Observational parameters}
\label{obs}
\begin{tabular}[h]{@{}llllllll@{}}
\hline
Rest frequency&Wavelength &FWHP& FWHP\footnote{At the distance of A\,1367 of 92 Mpc}&Receivers&Filterbank&$T_\mathrm{sys}$ \footnote{Typical $T_\mathrm{sys}$ ($T_\mathrm{mb}$ scale)}&$T_\mathrm{mb}$\footnote{Conversion to main beam brightness $T_\mathrm{mb}$= \textcolor{black}{$T^*_{A}$  $\eta_\mathit{fs}$ / $\eta_{mb}$} where $T_\mathrm{mb}$ is the main beam temperature,  \textcolor{black}{$\eta_\mathit{fs}$} is the forward scattering and spillover efficiency, and \textcolor{black}{$\eta_{mb}$} is the aperture efficiency. With \textcolor{black}{$\eta_\mathit{fs}$ / $\eta_{mb}$} from the IRAM website http://www.iram.fr. The main-beam temperatures can be converted into flux densities by applying a factor of \textcolor{black}{4.95} Jy/K. } \\
$[$GHz] & [mm] &[arcsec]&[kpc] &&[MHz]&[K]&[K]\\
\hline\\

115.3 &2.6 &22    &8.8 &A100, B100 &1&239&1.28 $T_{A}$*   \\
230.5 &1.3 &11    &4.4 &A230, B230 &4&679&1.70 $T_{A}$* \\
\hline
\end{tabular}
\end{minipage}
\end{table*}

A list of the observed galaxies and their basic characteristics is given in Table \ref{pramss}. A single CO observation was made at the optical centre (central pointing) of each of the 19 observed galaxies. Additionally for eight of the galaxies which had interaction signatures, e.g. disturbed morphologies at other wavelengths, we tried to determine, within the limits of sensitivity and pointing errors, whether their CO distribution was asymmetric or not. To do this we made maps with 3 to 6 pointings, typically with offsets along the optical major axis, in steps of 11 arcsec (half the 2.6 mm beam).

\begin{table*}
\centering
\begin{minipage}{140mm}
\caption{General characteristics of the observed galaxies}
\label{pramss}
\begin{tabular}[h]{lllllllrrrll}
\hline
Galaxy\footnote{Identifier from the Zwicky catalogue. The letters in parenthesis indicate relative position of the pair member. The equivalents in the NASA Extragalactic database (NED) are, \cg092(S) = \cg092(NED 01), \cg102(N) = \cg102(NED 02), \cg111(S) = \cg111(NED 01), and  \cg129(W) = \cg129(NED 01). } 	&	RA (2000)\footnote{From NED.} 	&	Dec (2000)$^b$	&	Type\footnote{Hubble type from GOLDMine \citep[Galaxy Online Database Milano Network,][http://goldmine.mib.infn.it/]{gava03b} or if unavailable NED.}	&	 Velocity\footnote{Optical velocity from NED.}	&	Major \footnote{Optical major axis from GOLDMine or if unavailable NED.}&	Minor&	$H_T$\footnote{Total extrapolated $H$--band  magnitude from GOLDMine.}&$V_{HI}$\footnote{\hi\ velocity from AGES with the exception of \cg079 which is from Paper I. }&$W_{20}$ \footnote{\hi\ velocity range, $W_{20}$ from AGES except for \cg079 which is taken from Paper I. }&State\footnote{Evolutionary state from Paper I.}&EW\footnote{EW(\halpha\ +[NII]) from GOLDMine.}  \\
&&&&&axis&axis&&&&&(\halpha+[NII])\\
CGCG	&	 [$^h$ $^m$ $^s$] 	&	 [\degree\ \prim\ \prin\ ] 	&	 	&	[\km ]	&	[\prim]&	[\prim]	&	[mag ]	&	[\km ]&	[\km ]& &\AA	\\
\hline\\	
97-062	&	11 42 14.55 	&	19 58 33.61 	&	Pec	&	7815	&	1.01	&	0.40	&	12.95	&	7774	&	246&A&\,\,37\\

97-063	&	11 42 15.70 	&	20 02 55.16	&	Pec	&	6102	&	0.58	&	0.34	&	13.56	&	6087	&	173&A&\,\,22\\

97-068	&	11 42 24.48 	&	20 07 09.90  	&	Sbc	&	5974	&	1.23	&	0.76	&	11.26	&	5979	&	353&B&\,\,41\\

97-072	&	11 42 45.20  	&	20 01 56.60  	&	Sa	&	6332	&	1.21	&	0.54	&	11.36	&	6338	&	307&B&\,\,\,\,9\\

97-076	&	11 43 02.17 	&	19 38 59.87   	&	Sb	&	6987	&	1.20	&	0.54	&	11.39	&	-	&&D&\,\,\,\,1	\\

97-079	&	11 43 13.40 	&	 20 00 17.00	&	Pec	&	7000	&	0.75	&	0.45	&	13.15	&	7073	&	238&A&129\\

97-078	&	11 43 16.22 	&	19 44 56.18	&	Sa	&	7542	&	1.88	&	0.92	&	11.22	&	-	&-&D&\,\,\,\,\,-	\\

97-082	&	11 43 24.49	&	19 44 59.95	&	Sa	&	6100	&	1.30	&	0.64	&	10.50	&	-	&-&D&\,\,\,\,-	\\

97-092(S)	&	11 43 58.20 	&	20 11 08.00 	&	Sbc	&	6487	&	0.76	&	0.54	&	12.57	&	-	&-&B&\,\,\,28	\\

97-091	&	11 43 59.00 	&	 20 04 37.00	&	 Sa	&	7368	&	1.12	&	0.81	&	11.07	&	7372	&	273&B&\,\,\,23\\

97-093	&	11 44 01.77	&	19 47 03.54	&	Pec	&	4909	&	0.96	&	0.39	&	12.90	&	-	&-&D&\,\,\,\,9	\\

97-102(N)	&	11 44 17.20  	&	20 13 23.90  	&	Sa	&	6368	&	1.08	&	0.65	&	11.40	&	6370	&	325&C&\,\,\,\,2\\

97-111(S)	&	11 44 25.91	&	20 06 09.70   	&	Irr	&	7436	&	0.45	&	0.30	&	12.77	&	-	&-&C&\,\,\,\,-	\\

97-121	&	11 44 47.06 	&	20 07 29.10  	&	Sab	&	6571	&	1.20	&	0.83	&	10.64	&	6583	&	408&C&\,\,\,\,4\\

97-114	&	11 44 47.80 	&	19 46 24.31 	&	Pec	&	8293	&	0.54	&	0.49	&	13.19	&	-	&-&-&\,\,36	\\

97-120	&	11 44 49.16 	&	19 47 42.14 	&	Sa	&	5595	&	1.32	&	0.85	&	10.55	&	-	&-&D&\,\,\,\,4	\\

97-122	&	11 44 52.23 	&	19 27 15.12 	&	Pec	&	5468	&	0.47	&	0.47	&	11.74	&	5478	&	456&B&\,\,45\\

97-125	&	11 44 54.85 	&	19 46 35.18 	&	S0a	&	8271	&	0.84	&	0.59	&	11.53	&	8225	&	425&-&\,\,23\\

97129(W)	&	11 45 03.91 	&	19 58 25.51 	&	Sb	&	5085	&	2.36	&	1.27	&	9.77	&	5091	&	494&C&\,14\\
\hline
\end{tabular}
\end{minipage}
\end{table*}

The data reduction was carried out with the CLASS software package (Forveille 1990). All scans were reviewed and those with poor baselines or other abnormalities were discarded. To measure line fluxes we defined a velocity window in which the line
was either detected or, in the case of non--detections,  was expected to be detected based on the \hi\ velocity width.
This window was typically centred on the optical velocity and was 150 -- 300\,\km\ wide.
The use of wobbler switching produced a flat baseline for each spectrum, allowing a 
linear fit to the two regions of the spectrum adjacent to the window which was then subtracted from the raw spectrum to produce the final spectrum.

\section{OBSERVATIONAL RESULTS}
\label{results}
The positions of the observed galaxies are shown in Figure \ref{coall} and details of all the observed spectra are tabulated in  Table \ref{cosum}, together with definitions of the properties derived from measurements of the spectra.  The central pointing  \hoz\ and \hto\ spectra for each galaxy are shown in Figures \ref{groups} and \ref{groups2}. In these figures the \hto\ spectra have been scaled so that their peak main beam temperature $T_\mathrm{mb}$ is equal to the \hoz\ maximum. The scaling factor to achieve this is shown at the top right above each spectrum.   Where there is an \hi\ detection  from AGES\footnote{Arecibo Galaxy Environment Survey \citep{cort08}} or the NRAO VLA for the galaxy its velocity and velocity width, \textcolor{black}{$W_{20}$},  are  indicated with a bar below these spectra (purple).

\footnotesize

\begin{onecolumn}
\begin{landscape}
\begin{center}
\begin{longtable}{@{}lrrrrrrrrrrrr@{}}
\caption{CO observations} \label{cosum} \\
\cline{1-12}
&           &  &       J = 1 $\rightarrow$ 0 & &&&&J = 2 $\rightarrow$ 1 \\
\cline{4-7} \cline{9-12}
Galaxy	&$\Delta$RA$ ^a$	&$\Delta$Dec$^a$&rms$^b$	&I(CO)$^{c,d} $		&$V^e$	&W$_{20}^f$	&	&rms$^b$	&I(CO)$^{c,d} $	&$V^e$	&W$_{20}^f$ \\

CGCG	&[\prin]	&[\prin]	&[mK]&[K\,\km]	&[\km]	&[\km]	&&[mK]&[K\,\km]	&[\km]	&[\km]	\\

\endfirsthead
\cline{1-12}
\vspace{0.01mm}\\
\noindent
97062	&	0	&	0	&	2.5	&	1.26$\pm$ 0.13	&	7832$\pm$\hspace{1.5mm}42	&	224$\pm$\hspace{1.5mm}84	&		&	5.6	&	2.28$\pm$ 0.28	&	7796$\pm$\hspace{1.5mm}27	&	171$\pm$\hspace{1.5mm}54\\
97062	&	-7.8	&	-7.8	&	2.8	&	0.54$\pm$ 0.14	&	7731$\pm$\hspace{1.5mm}21	&	\hspace{1.5mm}86$\pm$\hspace{1.5mm}42	&		&	4.5	&	0.79$\pm$ 0.17	&	7807$\pm$\hspace{1.5mm}53	&	192$\pm$106\\
97062	&	7.8	&	7.8	&	3.6	&	$<$0.38$\pm$ 0.14	&	-	&	-	&		&	5.6	&	0.98$\pm$ 0.23	&	7879$\pm$\hspace{1.5mm}13	&	\hspace{1.5mm}99$\pm$\hspace{1.5mm}21\\
97063	&	0	&	0	&	3.1	&	$<$0.43$\pm$ 0.16	&	-	&	-	&		&	9.4	&	$<$1.37$\pm$ 0.48	&	-	&	- \\
97068	&	0	&	0	&	3.8	&	13.72$\pm$ 0.25	&	5970$\pm$\hspace{2.5mm}4	&	350$\pm$\hspace{2.5mm}8 	&		&	7.7	&	20.52$\pm$ 0.50	&	 5976$\pm$\hspace{1.5mm}12 	&	297$\pm$\hspace{1.5mm}24\\
97068	&	0	&	11	&	3.2	&	7.17$\pm$ 0.2	&	5959$\pm$\hspace{1.5mm}19 	&	329$\pm$\hspace{1.5mm}38	&		&	6.9	&	3.77$\pm$ 0.45	&	5992$\pm$\hspace{1.5mm}85 	&	329$\pm$170 \\
97068	&	0	&	-11	&	3.2	&	10.29$\pm$ 0.2	&	 5986$\pm$\hspace{2.5mm}3 	&	318$\pm$\hspace{2.5mm}6 	&		&	8.4	&	9.9$\pm$ 0.53	&	 5976$\pm$\hspace{1.5mm}23	&	 339$\pm$\hspace{1.5mm}46 \\
97068	&	0	&	-22	&	3.2	&	1.14$\pm$ 0.21	&	 5965$\pm$\hspace{1.5mm}81	&	 318$\pm$162	&		&	10.5	&	$<$1.48 $\pm$ 0.59	&	 - 	&	 - \\
97068	&	0	&	22	&	3.1	&	1.11$\pm$ 0.18	&	 5991$\pm$\hspace{1.5mm}67	&	329$\pm$134	&		&	6.6	&	$<$0.03$\pm$ 0.37	&	 - 	&	 - \\
97072	&	0	&	0	&	2.9	&	4.16$\pm$ 0.18	&	 6317$\pm$\hspace{1.5mm}26	&	287$\pm$\hspace{1.5mm}52	&		&	6.0	&	4.51$\pm$ 0.37	&	 6308$\pm$\hspace{1.5mm}37 	&	 266$\pm$\hspace{1.5mm}74 \\
97072	&	-8.9	&	6.5	&	3.6	&	3.65$\pm$ 0.23	&	 6334$\pm$\hspace{1.5mm}38	&	276$\pm$\hspace{1.5mm}76	&		&	7.2	&	2.22$\pm$ 0.47	&	 6345$\pm$\hspace{1.5mm}52 	&	234$\pm$104\\
97072	&	8.9	&	-6.5	&	2.0	&	2.42$\pm$ 0.12	&	 6253$\pm$\hspace{1.5mm}12	&	159$\pm$\hspace{1.5mm}24 	&		&	5.2	&	1.52$\pm$ 0.27	&	 6223$\pm$\hspace{1.5mm}18 	&	138$\pm$\hspace{1.5mm}36 \\
97076	&	0	&	0	&	2.1	&	1.63$\pm$ 0.14	&	 6993$\pm$\hspace{1.5mm}36	&	373$\pm$\hspace{1.5mm}72 	&		&	6.2	&	1.75$\pm$ 0.40	&	 6899$\pm$\hspace{1.5mm}29 	&	181$\pm$\hspace{1.5mm}58 \\
97079	&	0	&	0	&	3.0	&	1.29$\pm$ 0.15	&	 6985$\pm$\hspace{1.5mm}26	&	159$\pm$\hspace{1.5mm}52 	&		&	6.1	&	2.36$\pm$ 0.30	&	 7008$\pm$\hspace{1.5mm}42	&	202$\pm$\hspace{1.5mm}84 \\
97079	&	-7.8	&	7.8	&	2.6	&	1.31$\pm$ 0.13	&	 7033$\pm$\hspace{1.5mm}17	&	149$\pm$\hspace{1.5mm}34 	&		&	6.1	&	2.32$\pm$ 0.31	&	 7029$\pm$\hspace{1.5mm}51	&	202$\pm$102 \\
97079	&	-11	&	11	&	3.6	&	1.66$\pm$ 0.18	&	7044$\pm$\hspace{1.5mm}27 	&	170$\pm$\hspace{1.5mm}54 	&		&	6.5	&	0.91$\pm$ 0.33	&	7008$\pm$\hspace{2.5mm}3 	&	 \hspace{1.mm}26$\pm$\hspace{2.5mm}9\\
97079	&	-15.6	&	15.6	&	3.7	&	1.36$\pm$ 0.18	&	 7033$\pm$\hspace{2.5mm}9 	&	149$\pm$\hspace{1.5mm}18	&		&	5.8	&	1.28$\pm$ 0.28	&	 7093$\pm$\hspace{1.5mm}58	&	 202$\pm$116 \\
97078	&	0	&	0	&	2.1	&	 $<$0.30$\pm$ 0.12	&	 - 	&	 - 	&		&	4.9	&	 $<$0.70$\pm$ 0.28	&	 - 	&	 - \\
97082	&	0	&	0	&	2.4	&	 $<$0.34$\pm$ 0.14	&	 - 	&	 - 	&		&	5.8	&	 $<$0.81$\pm$ 0.32	&	 - 	&	 - \\
97091	&	0	&	0	&	2.6	&	7.18$\pm$ 0.15	&	 7385$\pm$\hspace{1.5mm}12	&	 267$\pm$\hspace{1.5mm}24 	&		&	4.1	&	12.51$\pm$ 0.22	&	 7381$\pm$\hspace{1.5mm}26 	&	 235$\pm$\hspace{1.5mm}52 \\
97092(S) 	&	0	&	0	&	3.8	&	3.37$\pm$ 0.19	&	 6472$\pm$\hspace{2.5mm}5 	&	159$\pm$\hspace{1.5mm}10 	&		&	6.7	&	5.71$\pm$ 0.33	&	 6473$\pm$\hspace{2.5mm}7 	&	 159$\pm$\hspace{1.5mm}14 \\
97092(S) 	&	0	&	-11	&	2.8	&	2.1$\pm$ 0.12	&	 6493$\pm$\hspace{2.5mm}2	&	117$\pm$\hspace{2.5mm}4 	&		&	7.4	&	1.54$\pm$ 0.31	&	 6495$\pm$\hspace{2.5mm}5 	&	\hspace{1.5mm}74$\pm$\hspace{1.5mm}10 \\
97092(S) 	&	0	&	11	&	3.3	&	2.06$\pm$ 0.15	&	 6483$\pm$\hspace{1.5mm}12	&	 138$\pm$\hspace{1.5mm}24 	&		&	6.3	&	1.78$\pm$ 0.29	&	 6452$\pm$\hspace{1.5mm}14	&	 117$\pm$\hspace{1.5mm}28 \\
97092(S) 	&	11	&	0	&	4.4	&	1.65$\pm$ 0.2	&	 6493$\pm$\hspace{1.5mm}32	&	159$\pm$\hspace{1.5mm}64 	&		&	7.0	&	$<$0.98$\pm$ 0.39	&	 - 	&	 - \\
97092(S) 	&	-11	&	0	&	4.0	&	2.38$\pm$ 0.19	&	 6472$\pm$\hspace{2.5mm}8	&	159$\pm$\hspace{1.5mm}16 	&		&	7.6	&	2.72$\pm$ 0.35	&	 6473$\pm$\hspace{1.5mm}43 	&	 159$\pm$\hspace{1.5mm}86 \\
97092(S) 	&	0	&	22	&	4.6	&	0.99$\pm$ 0.21	&	 6430$\pm$\hspace{1.5mm}14	&	 \hspace{1.5mm}74$\pm$\hspace{1.5mm}28 	&		&	11.4	&	$<$1.61$\pm$ 0.64	&	 - 	&	 - \\
97093	&	0	&	0	&	2.5	&	1.35$\pm$ 0.15	&	 4873$\pm$\hspace{1.5mm}28	&	201$\pm$\hspace{1.5mm}56 	&		&	6.4	&	2.3$\pm$ 0.33	&	 4895$\pm$\hspace{1.5mm}41 	&	 201$\pm$\hspace{1.5mm}82 \\
97102(N) 	&	0	&	0	&	2.8	&	2.35$\pm$ 0.18	&	 6369$\pm$\hspace{1.5mm}53	&	 297$\pm$106 	&		&	7.0	&	2.62$\pm$ 0.43	&	 6381$\pm$\hspace{1.5mm}78 	&	 276$ \pm $156 \\
97102(N) 	&	7.2	&	-8.3	&	4.6	&	1.72$\pm$ 0.26	&	 6491$\pm$\hspace{2.5 mm}9	&	\hspace{1.5mm}95$\pm$\hspace{1.5mm}18 	&		&	10.5	&	$<$1.48$\pm$ 0.59	&	 - 	&	 - \\
97102(N) 	&	-7.2	&	8.3	&	3.3	&	2.61$\pm$ 0.21	&	 6369$\pm$\hspace{1.5mm}42	&	 297$\pm$\hspace{1.5mm}84 	&		&	7.3	&	2.31$\pm$ 0.45	&	 6391$\pm$\hspace{1.5mm}53 	&	 255$\pm$106 \\
97111(S) 	&	0	&	0	&	2.5	&	$<$0.35$\pm$ 0.14	&	 - 	&	 - 	&		&	7.3	&	$<$1.0$\pm$ 0.40	&	 - 	&	 - \\
97114	&	0	&	0	&	3.4	&	1.94$\pm$ 0.16	&	 8481$\pm$\hspace{1.5mm}15	&	 139$\pm$\hspace{1.5mm}30 	&		&	6.3	&	4.29$\pm$ 0.29	&	 8461$ \pm$\hspace{2.5mm}8 	&	139$\pm$\hspace{1.5mm}16\\
97120	&	0	&	0	&	3.2	&	8.66$\pm$ 0.22	&	 5622$ \pm$\hspace{1.5mm}23	&	413$\pm$\hspace{1.5mm}46 	&		&	4.7	&	11.32$\pm$ 0.32	&	 5571$\pm$\hspace{1.5mm}25 	&	 308$\pm$\hspace{1.5mm}50 \\
97120	&	-6.8	&	-8.7	&	5.6	&	7.69$\pm$ 0.38	&	 5633$\pm$\hspace{1.5mm}32	&	 392$\pm$\hspace{1.5mm}64 	&		&	10.7	&	4.38$\pm$ 0.56	&	 5762$\pm$\hspace{1.5mm}23 	&	 138$\pm$\hspace{1.5mm}46 \\
97120	&	6.8	&	8.7	&	5.5	&	6.9$\pm$ 0.38	&	 5495$\pm$\hspace{1.5mm}15	&	 159$\pm$\hspace{1.5mm}30 	&		&	10.3	&	5.87$\pm$ 0.71	&	 5454$\pm$\hspace{2.5mm}9 	&	\hspace{1.5mm}75$\pm$\hspace{1.5mm}18 \\
97120	&	9.2	&	11.8	&	9.1	&	5.24$\pm$ 0.63	&	 5464$\pm$\hspace{1.5mm}15	&	 \hspace{1.5mm}96$\pm$\hspace{1.5mm}30 	&		&	15.9	&	$<$2.24$\pm$ 0.90	&	 - 	&	 -\\
97121	&	0	&	0	&	2.4	&	1.58$\pm$ 0.17	&	 6572$ \pm$\hspace{1.5mm}77	&	383$\pm$154 	&		&	4.4	&	2.19$\pm$ 0.30	&	 6584$\pm$\hspace{1.5mm}96 	&	 404$\pm$196\\
97122	&	0	&	0	&	3.1	&	6.67$\pm$ 0.21	&	 5522$\pm$\hspace{1.5mm}19	&	 360$\pm$\hspace{1.5mm}38 	&		&	4.7	&	9.67$\pm$ 0.30	&	 5502$\pm$\hspace{1.5mm}11 	&	 233$\pm$\hspace{1.5mm}22 \\
97125	&	0	&	0	&	2.1	&	2.85$\pm$ 0.16	&	 8224$\pm$\hspace{1.5mm}72	&	 481$\pm$144 	&		&	4.8	&	5.92$\pm$ 0.36	&	 8236$\pm$\hspace{1.5mm}70 	&	 502$\pm$140\\
97129(W)	&	0	&	0	&	3.9	&	4.05$\pm$ 0.3	&	 5134$\pm$\hspace{1.5mm}89	&	 412$\pm$178 	&		&	8.7	&	1.97$\pm$ 0.57	&	5108$\pm$143 	&	 381$\pm$286\\
97129(W)	&	-2.3	&	-10.8	&	3.7	&	2.74$\pm$ 0.27	&	 5044$\pm$\hspace{1.5mm}77	&	 380$\pm$154 	&		&	8.1	&	2.17$\pm$ 0.46	&	 5098$\pm$120 	&	402$\pm$240\\
97129(W)	&	2.3	&	10.8	&	4.5	&	3.13$\pm$ 0.36	&	 5155$\pm$110	&	455$\pm$220 	&		&	7.5	&	2.9$\pm$ 0.38	&	 5172$\pm$\hspace{1.5mm}54 	&	 254$\pm$108\\
\cline{1-12}
\end{longtable}
\end{center}
\end{landscape}
\end{onecolumn}

\begin{twocolumn}
\noindent 
\textbf{Table \ref{cosum} notes}\\
({\em $a$}) Pointing offset from the galaxy centre in arcsec.\\
({\em $b$}) rms noise; channel width (\textcolor{black}{$\delta V_{CO}$ }) was 10.6 \km.\\
({\em $c$}) \ico= $\sum_i$ $T^i_\mathrm{mb}\Delta V$ with the summation carried out across a velocity window determined from the range of velocities over which emission was observed. For non--detections we give upper limits based on 2.5 times the rms within the \textcolor{black}{velocity} window \textcolor{black}{(i.e. $\delta$\ico, see below)}.  \\ 
({\em $d$}) $\delta$ \ico  =$\sigma \Delta V N^{1/2}$, where \textcolor{black}{$\delta V_{CO}$ } = channel velocity width -- 10\,\km, $\sigma$ = rms brightness temperature (in channels of \textcolor{black}{$\delta V_{CO}$ } width), $N$ = number of channels within the velocity window.\\
({\em $e$}) $V$= the mean  of the lower and upper $W_{20}$ velocity limits from the  spectrum smoothed to 20\,\km. The uncertainty, $ \sigma (V)= (W_{20} - W_{50}$)(S/N)$^{-1}$ is based on \cite{schnei90}, except where $W_{20} = W_{50}$ in which case $\sigma$(V) =  (2 $\times$  \textcolor{black}{$\delta V_{CO}$ }) (S/N)$^{-1}$. W$_{20}$ and W$_{50}$  are the line widths at 20\% and 50\% of the peak intensity. The signal to noise (S/N) is defined as the spectrum's peak flux  over the rms noise. In the case of marginal detections (S/N $\sim$ 2.5--3)  the mean velocity was determined from the mean of a single Gaussian fit to the emission within the velocity window. The uncertainty then is taken as the  1$\sigma$ uncertainty of the fit. \\ 
({\em $f$}) W$_{20}$ as defined in (e) above with W$_{20}$ uncertainty = 2 $\sigma$(V). The exception being the marginal detections where the velocity width and uncertainties are the FWHM and its uncertainty from the Gaussian fit referred to in note (e).\\

\normalsize
\begin{figure*}
\begin{center}
\includegraphics[ angle=0,scale=1.0] {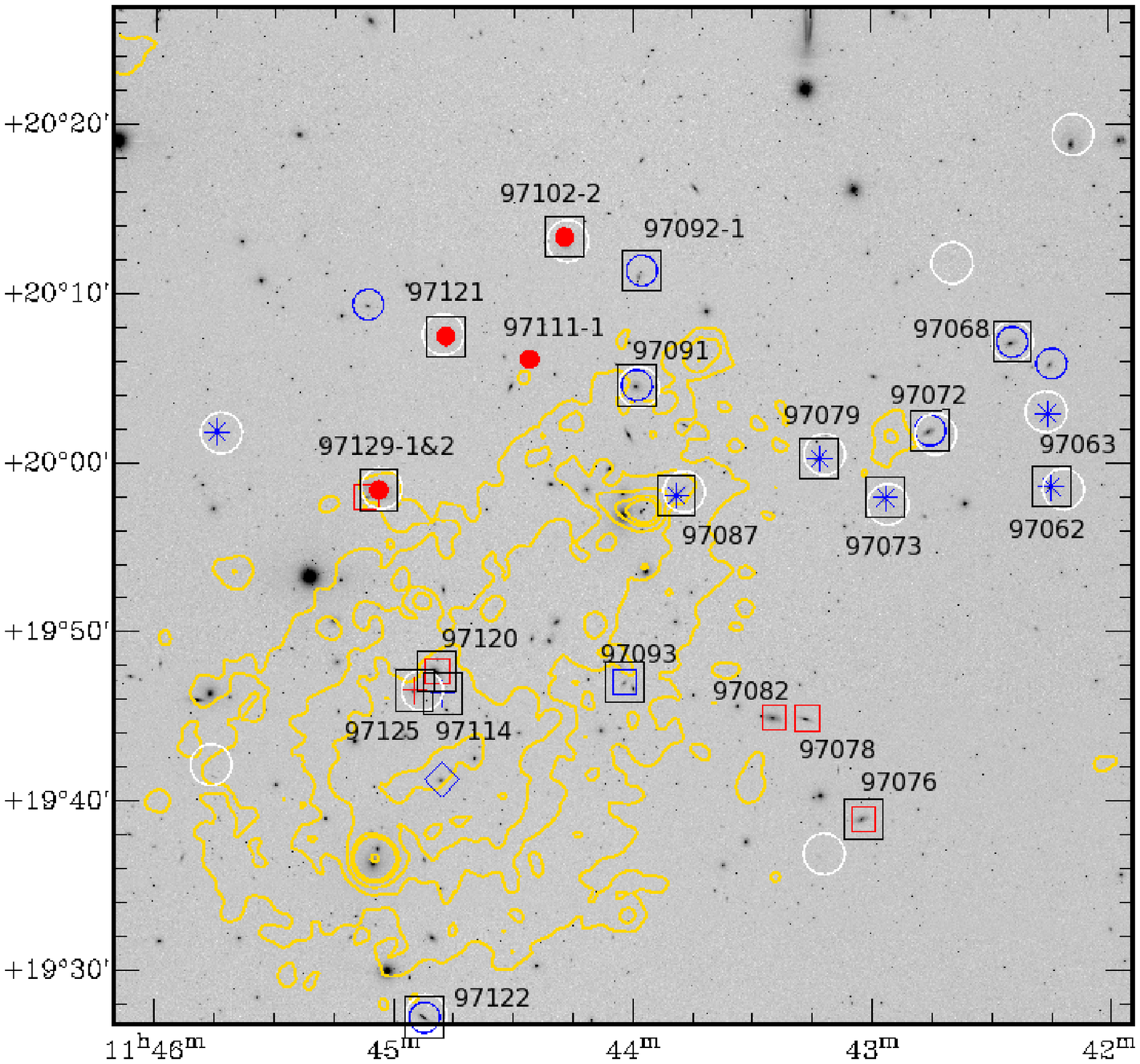}
\vspace{1cm}
\caption{Positions of bright late--type galaxies in the central region of A\,1367 from Paper I. Our CO detections plus \cg073 and \cg087 from \citep{bosel94} are shown with black open squares. The  \emph{colour} of the symbol (red or blue) used for each spiral indicates its SDSS \textit{g--i} colour, blue ($\leq$1.1) red ($>$1.1). AGES \hi\ detections are marked (white circles). The \emph{symbol} used for each spiral indicates its evolutionary state from Paper I  as follows; A--(asterisk), B--(open circle), C--(filled circle) and D--(open square). Spirals that belong to BIG are marked with a cross and unclassified spirals with a diamond.  {\em ROSAT} X--ray intensities from archive data (yellow contours) are overlaid on an SDSS \textit{i}--band image. Details for the galaxy
identifiers are given in note (a) of Table. \ref{pramss}  }
\label{coall}
\end{center}
\end{figure*}

\begin{figure*}
\begin{center}
\includegraphics[ angle=0,scale=0.8] {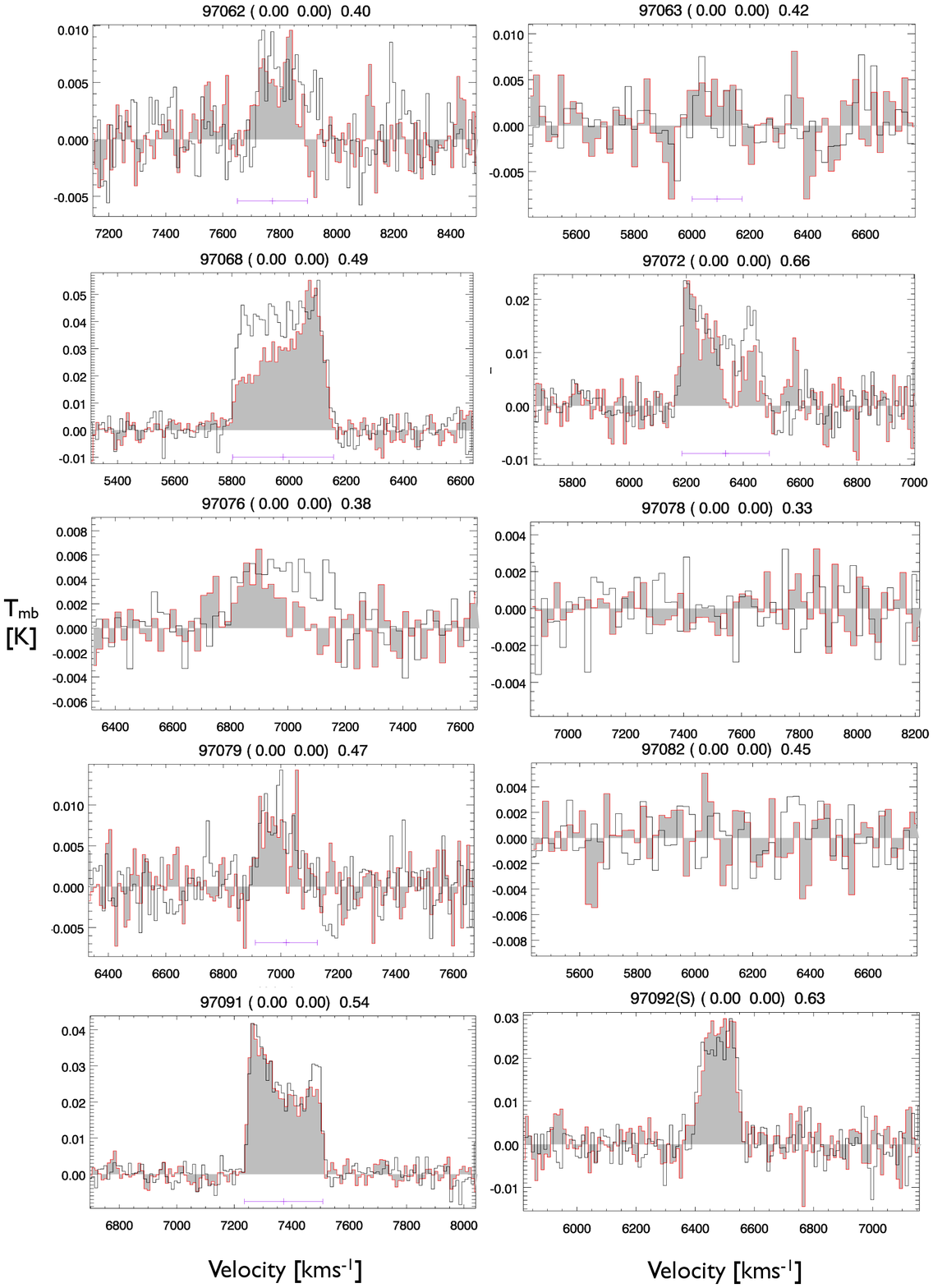}
\vspace{1cm}
\caption{\hoz\ -- black  and  \hto\ -- red spectra  at the position of the optical centre of each galaxy.  $\delta V_{CO}$ = 10\,\km, except \cg063, \cg076, \cg078, \cg082 and \cg092(S) where it is 20\,\km. The \hto\ spectra are scaled so that the \hto\ $T_\mathrm{mb}$ maximum has the same value as the \hoz\ maximum;  the  scaling factor is shown at the top right above the spectrum.  The vertical scale is $T_\mathrm{mb}$ in Kelvin and the horizontal scale is the velocity in \km.  Where there is an \hi\ detection (AGES or VLA Paper I) for the galaxy its central velocity and W$_{20}$ velocity width is shown below the spectra (purple). }
\label{groups}
\end{center}
\end{figure*}

\begin{figure*}
\begin{center}
\includegraphics[ angle=0,scale=0.80] {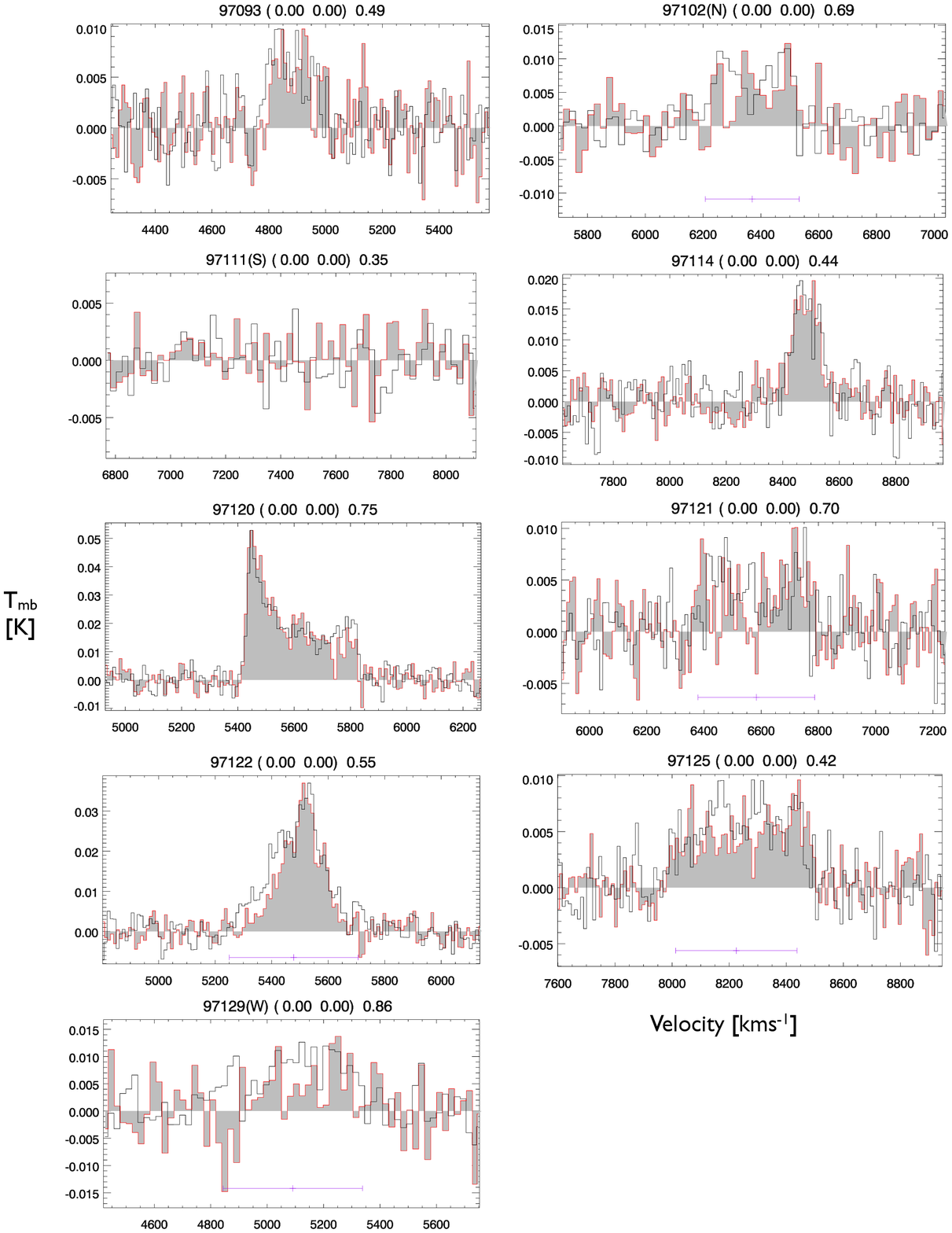}
\vspace{1cm}
\caption{\hoz\ -- black  and  \hto\ -- red spectra  at the position of the optical centre of each galaxy.  $\delta V_{CO}$ = 10\,\km, except \cg102(N), \cg111(S) and \cg129(W) where it is 20\,\km. Further detail for these spectra are given in the caption to Figure \ref{groups}. }
\label{groups2}
\end{center}
\end{figure*}

For eight of the spirals,  \cg062, \cg068, \cg072,  \cg079 \cg092(S), \cg102(N), \cg120 and \cg129(W), sample maps were made with 3 to 6 pointings. The intensities and pointing offsets for these spirals, relative to the central pointing, are given in columns 2 and 3 of Table 3.   Appendix \ref{amp}, Figures \ref{offs62} to \ref{offs129}, show the spectra and pointing positions for these spirals.  Where, for one of the emission lines, there were more than three detections projected linearly across the spiral we fitted  a Gaussian to the intensity distribution as a function of projected position. The Gaussian fits for  \cg079 and  \cg102(N) produced significant indications of asymmetric molecular gas distributions (See section \ref{maps} and Appendix \ref{amp}).

\section{Discussion}
\label{discuss}

Section \ref{mdef}  considers whether the observed spirals are deficient in molecular hydrogen, based on the \hoz\ intensities.  The   \hdos\ and \hi\ deficiencies of the observed spirals are considered in section \ref{coresult} while their star formation rates relative to their \hdos\ reservoirs are analysed in section \ref{sfrs}. In  section \ref{maps} we  assess,  for the subset of the observed spirals with multiple pointings, what evidence the pointings provide about the distribution of the molecular gas in those spirals.

\subsection{H$_2$ deficiency}
\label{mdef}

We aim to determine whether the observed spirals are \hdos\ deficient or not. A spiral's \hdos\ deficiency (\hd)  is defined in  \cite{bosel97} as:
\begin{equation}
\hspace{7 mm} \mathit{Def}_{H_2} = \log_{10}(M_{H_2})_e - \log_{10}(M_{H_2})_o    \\
\end{equation}
where  \me\ is  the expected \hdos\ mass and  \mo\ is the observed \hdos\ mass. 

It is critical in deriving the H$_2$ deficiency of a spiral that \me\ is determined from a galaxy characteristic which is as independent as possible from its current molecular content, but is strongly correlated  to the  $M_{H_2}$ contained in isolated spirals of similar Hubble type and size.  Using a sample of optically selected non--\hi\ deficient spirals and the standard conversion factor (also known as the $X$--factor) between \ico\ and \hdos, \cite{bosel97} determined  such a relation between  $M_{H_2}$ and $L_H$ (and alternatively with linear size) with no significant residuals attributable to Hubble type. However the good correlation between a spiral's  total $L_H$ and its dynamical mass \citep{gava96,bosel01}, implies we are relying on  a more fundamental relation between a spiral's \me\ and its dynamical mass. Subsequently  \cite{bosel02}  modified the relation to determine $M_{H_2}$  derived from a $L_H$ dependent conversion factor, again using an optically selected sample of \hi\ non--deficient spirals (\defhi\   $\leq 0.3$). Their sample galaxies were essentially the same as those used by   \cite{hayn84}  to calibrate \hi\ deficiencies.

We have used the  \cite{bosel02} relation (their equation 8)  to calculate the expected \hdos\ mass, \me, i.e. 
\begin{equation}
\hspace{7 mm} \log_{10}(\mathrm{\textcolor{black}{\textit{M}}_{H_2})_e} =3.28+0.51 \log_{10}(L_H)   \\
\end{equation}
\begin{equation}
\hspace{7 mm} \log_{10} (L_H) = 11.36 -0.4H_T+2 \log_{10}D \\
\end{equation}

where $H_T$ = the galaxy's total \textit{H}--band magnitude \citep{gava00}, $D$ = distance in Mpc  and $M_{H_2}$ and $L_H$ are expressed in solar units. 

Each spiral's \textit{actual} \mo\ was  determined, from its observed  \ico, which incorporates a luminosity dependent $X$--factor from \cite{bosel02},  and assuming the source was unresolved in the \hoz\ beam (see Appendix \ref{cofm})  i.e: 
\begin{equation}
\hspace{7 mm} \mathrm{M}(H_2)_o =7950\,\chi_{co}\, D^2\, \sum_i S_i \Delta V
\end{equation}

\begin{equation}
\hspace{7 mm} \sum_i S_i \Delta V = I(CO) \times G \\
\end{equation}
 \textcolor{black}{
where:\
\begin{description}
\itemsep-0.3cm
\item the coefficient 7950 is derived in Appendix \ref{cofm}; \\
\item \textcolor{black}{$G$} = 4.95\,Jy\,K$^{-1}$ is the telescope gain derived in Appendix \ref{cofm};\\
\item $\sum_i S_i \Delta V$ = flux in Jy\,km\,s$^{-1}$; \\ 
\item $\chi_{co}$ = $X_{CO}^H$ / $X_{CO}$, where $X_{CO}^H$ is the $L_H$ dependent conversion factor from Table 5 in  \cite{bosel02}, i.e. $\log_{10}(X_{CO}^H)$ = -0.38 $\log_{10}(L_H)$ + 24.23 and $X_{CO}$  = 2.3 $\times 10^{20}$ mol cm$^{-2}$ (K\,\km)$^{-1}$ \citep{strong88};\\
\item$D$ = distance in Mpc; \\
\item \ico\ = CO intensity in K\,\km ($T_\mathrm{mb}$ scale);\\
\item $H_T$ = Total \textit{H}--band magnitude from GOLDMine \citep{gava03b}. 
\end{description} }
Our calculations of \me\  and \mo\  are both based on \textit{H}--band luminosity ($L_H$), rather than $L_B$, in order to minimise the impact of temporary interaction induced enhancements in  \LB\  and of extinction.

In calculating \mo\ we have assumed that the  \hoz\ emission was unresolved in the 2.6 mm IRAM beam. CO emission from a spiral's disk is generally understood to decline exponentially \citep{young95,leroy09} with radius from the optical centre. \textcolor{black}{ From a  sample of barred and unbarred spirals \cite{nishiyama01}  found a mean scale length  $r_e(CO)/r_{25}$ = 0.22 $\pm$0.12 ($r_e$ = 2.5 $\pm$1.6 kpc).} The FCRAO\footnote{Five College Radio Astronomy Observatory}  survey found a decline to 1 K \km\  (CO isophotal radius) at \aprox\ 52\% of the optical radius based on $D_{25}$   \citep{young95}. Typically our observed spirals have an optical major axis diameter of $\sim $ 60 arcsec (24.8\,kpc) implying CO isophotal radii of $\sim$ 6.2\,kpc, or diameters of $\sim$ 12.4\,kpc, i.e., somewhat larger than the IRAM 30--m beam at 2.6\,mm of 22\,arcsec (9\,kpc).  Ten of the spirals of our sample have been previously  observed in \hoz\ with the NRAO Kitt Peak 12--m telescope which has a 55 arcsec beam by \citet{bosel97} which means that we can check to what extent we might be underestimating their CO flux (see Appendix \ref{appc}). This analysis confirms that the \hoz\ emission fills a substantial fraction of the IRAM--30m beam and, in those cases where we only have a single pointing might be systematically underestimating their CO flux by up to 30\%. For some targets, in particular the largest ones, we might be off by up to a factor of two.

Where \hoz\ emission extends  beyond the (22\,arcsec)  full width half maximum (FWHM) of the IRAM 30--m beam at the central pointing position (large or extended objects) the \mo\ mass  shown in  Table \ref{calc} is a lower limit. In the case of \cg087 and \cg079 the \mo\ calculations include emission detected beyond the 2.6\,mm central pointing beam, i.e. from multiple pointings. \cg129W is a special case, this being  by far the largest observed spiral ($D_{25}$ =2.3 arcmin), nearly twice the optical size of the next largest observed spiral (see Table \ref{pramss}). \cg129W shows evidence that its CO is distributed across a large fraction of its optical disk (see Appendix \ref{appc}).  For this reason the \mo\ shown in  Table \ref{calc} for \cg129W has been derived from the Kitt Peak \hoz\ flux rather the IRAM 30--m \hoz\ flux. 

\subsection{H$_2$ content -- results}
\label{coresult}
The  \hdos\ deficiencies for our sample, determined from our \hoz\ pointings (in most cases from the central pointings alone), and   for \cg129W from the Kitt Peak pointing, are given in Table \ref{calc}. The mean and standard deviation in  \hd, for the detections in our sample and \cg073 and \cg087 from \cite{bosel94}    \citep[excluding spirals in the Blue Infalling Group -- BIG][]{cort06}    is  -0.26 $\pm$ 0.35.  32\% of the observed spirals have \hdos\ excesses (negative deficiencies)  $<$ -0.2  compared to 17 \% in the \cite{bosel97} sample of isolated galaxies (their Figure 7). Even larger \hdos\ excesses have been reported in spirals of a sample of less evolved Hickson Compact Groups \citep{badenes12} which was calibrated  to a different isolated galaxy sample \citep[AMIGA][]{lisenfeld11}. Further observations are required, mapping the full extent of the CO distribution, to determine to what extent these apparent surpluses in both the Martinez--Badenes and our own samples are genuine, for example due to the objects being perturbed and as a result of this the balance in the ISM being skewed to the formation of \hdos, or if they fall within the range of uncertainty that is typical in determining \hdos\ deficiencies.  
 
Figure \ref{defdist} shows box plots of the  \hdos\ deficiencies or lower limits for the observed spirals plus \cg073 and \cg087 in each of the four evolutionary states from Paper I and defined in terms of \hi\ deficiency and SDSS \textit{g -- i} colour in Table \ref{ttt}. Despite the uncertainties about the absolute \hdos\ deficiencies, which are discussed further in Appendix \ref{appc},  Figure \ref{defdist}  shows a   clear relative difference in the distribution of  \hdos\ deficiencies between spirals in less evolved evolutionary states,  A and B, and those with more advanced evolutionary states,  C  and D. This relative difference  is reflected in  the mean deficiencies  for the galaxies in each evolutionary state shown  in Table \ref{ttt}  as well as the log rank (Mantel -- Haenszel) test  for samples containing censored data (non--detections), which gives a probability of   the AB and CD samples coming from the same  population of only 0.04\% (\textcolor{black}{$p$} value = 0.0004). 

\begin{figure}
\centering
\includegraphics[scale=0.56] {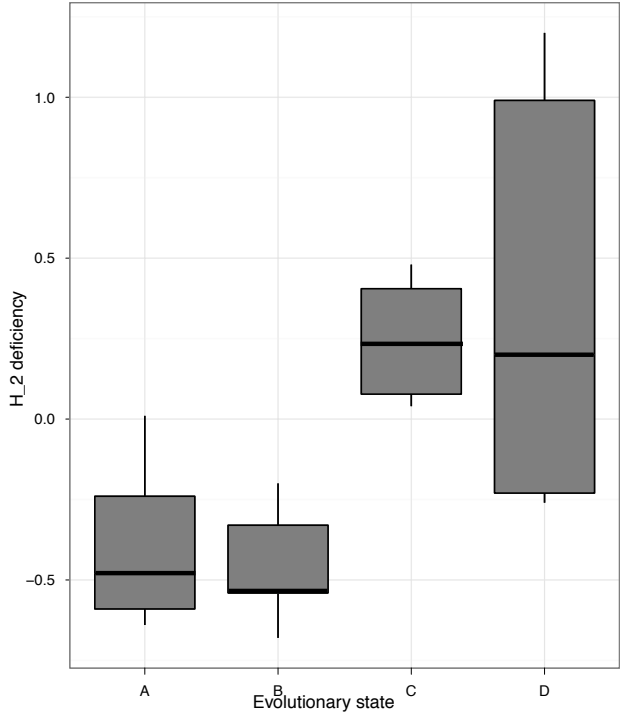}
\caption{  \hdos\  deficiencies for the observed spirals plus \cg087,  \cg073 in each evolutionary state,  including non--detections but excluding spirals in the Blue Infalling Group. The horizontal thicker black line within each box shows the median \hdos\  deficiency for the state with upper and lower edges of the grey boxes indicating the upper and lower quartiles. The ends of the vertical lines show the upper and lower range (inclusive of  lower limits) of  \hdos\ deficiency for spirals in each evolutionary state.}
\label{defdist}
\end{figure}

Figure \ref{defcomp} compares \hd\ and \dhi\ for each of the observed spirals  with symbols indicating their evolutionary state from Paper I (Table \ref{ttt}).  We used the estimates of natural variation in \dhi\ ($\pm$ 0.3 dex from Paper I) and \hd\ ($\pm$ 0.4 dex), indicated by dashed  vertical lines and horizontal lines respectively  in Figure \ref{defcomp}, to identify the following classes of spirals, defined in Table \ref{gasdef}: \hi\ deficient, \hdos\ deficient and gas deficient, i.e. spirals with both an  \hi\ and \hdos\ deficiency. \textcolor{black}{When interpreting the deficiencies in Figure \ref{defcomp} it is important to note that the value of the deficiency is subject to  measurement errors (e.g., beam filling factor, CO to \hdos\ conversion, observational errors in the reference sample). An additional uncertainty in assessing an individual  deficiency arises from the large natural variation in \hi\ and \hdos\ content  of spirals of the same morphological type in the reference field sample. This means that the further an \hi\ or \hdos\ deficiency value is from the deficiency/excess threshold (measurement and natural variation),} delineated by the dashed rectangle in the figure, the more confident we can be that it is truly deficient. This is the case irrespective of whether the deficiency is determined based on a detection or a lower limit from a non--detection.  Kendell's $\tau$ correlation coefficient, taking into account the fact that both deficiencies contain censored data,  for  \hdos\ and \hi\ deficiencies is 0.064,  indicating \hdos\ and \hi\ deficiencies are not correlated. But interestingly the same test for correlation between  \hdos\ deficiency and  evolutionary type has a  Kendell's $\tau$ of 0.27 which would reject the null hypothesis of no correlation at the 10\% level. The correlation between \hdos\ deficiency and evolutionary type (based on \textit{g--i} colour and \dhi) and lack of correlation with \dhi\ argues for a connection between \hdos\ deficiency and a process connected with colour evolution (possibly a burst of enhanced SF) and against direct removal of molecular gas by ram pressure stripping. However almost all of the points in Figure \ref{defcomp} are on or below the 1-to-1 correlation confirming that the \hdos\ is less affected by the environment than the \hi. Whatever the physical mechanism, this supports a scenario in which the \hdos\ is more tightly bound to the galaxy and less affected by external mechanisms.

The calculation of \mo\ is sensitive to distance errors, but as the distance also enters the calculation of \me\ via $L_H$ (Equation 2), \hd\ is much less sensitive to errors in $D$. Furthermore, by adopting the same conversion factor ($X$--factor) between \hoz\ and $M_{H_2}$	when determining both \me\ and \mo, means uncertainties attributable to the conversion factor are to a great extent cancelled out. The ratio between the luminosity--dependent and the	standard,	$2.3$  $\times$  10$^{20}$ mol cm$^{-2}$ (K\,\km)$^{-1}$  \citep{strong88,Polk88}, $X$--factors for the observed galaxies ranges from 0.35 to 1.32.

\begin{table*}
\centering
\begin{minipage}{140mm}
\caption{\hdos\ deficiencies from \hoz\ and star formation rates for the central pointings}
\label{calc}
\begin{tabular}[h]{@{}lrrrrrrrrr@{}}
\hline
&\hi\ &&&&\hdos \\

 \cline{2-3} \cline{5-8}  
Galaxy\footnote{The \cg073 and \cg087 \ico\ data are from \cite{bosel94} and marked in the table with an asterisk. }	&	M expected\footnote{from Paper I.}	&	Def\footnote{from Paper I.}		&&	M expected	&	M observed\footnote{$M_{H_2}$ is derived from our IRAM  \hoz\  except for  \cg129W which is derived from the Kitt Peak 12--m observations by \cite{bosel97}. The IRAM derived mass for \cg129W is 0.48 x 10$^9$ \msolar. }	&	\ico \footnote{ The \ico$_{Tmb}$ for \cg079 is the sum of the \hoz\ flux from the central pointing  plus that from the [-15.6,15.6] pointing, which was included because the flux it detected lies beyond the FWHM of the  \hoz\ central pointing beam. Similarly for \cg087 the flux is the sum of fluxes from the [-21,21] and central pointings from \cite{bosel94}. The \hoz\ detection for \cg063 was marginal and there were no detections for \cg078, \cg082, and \cg111(S) so  the values given are  upper limits (see section \ref{mdef}).  }	&Def\footnote{\hd is derived from our IRAM  \hoz\  observations except for  \cg129W which is derived from the Kitt Peak 12--m observations by \cite{bosel97}. The IRAM derived \hd\ for \cg129W is  0.4. See section \ref{mdef} for details of the method used to calculate  \hd.}	&SFR\footnote{SFR = SFR(\halpha)  using the method from \cite{bosel02}. Corrections (dependent on morphological type) were made for [NII] contamination and internal extinction as per \cite{bosel01}.  No adjustment has been made for gas recycling as the appropriate rate to use in the central cluster environment is unknown and it is thought any expelled gas returning to a galaxy would be unlikely to remain molecular.} &\hdos\ depletion\footnote{$M_{H_2}$/SFR(\halpha). $M_{H_2}$ is derived from our IRAM  \hoz\  except for  \cg129W which is derived from the Kitt Peak 12--m observations by \cite{bosel97}. The $M_{H_2}$/SFR(\halpha) for \cg129W derived using he IRAM \hoz\ observations is 0.3 Gyr.  } \\
&&&&&&&&&timescale\\
CGCG	&	[10$^9$ \msolar ] 	&		&&	[10$^9$ \msolar ]	&	10$^9$ [ \msolar]	&	[K\,\km ]	&	&[ \msolar  yr$^{-1}$ ]	&[Gyr]\\

\hline 
97062	&3.93	&0.35	&	&0.27	&0.47	&1.26		&	-0.22		&0.32 	&1.48\\
97063	&1.50	&0.01	&	&0.20	&$\leq$0.20	&$\leq$0.43	&	$\geq$0.01	&0.16	&1.23\\
97068	&4.48	&-0.28	&	&0.60	&2.87	&13.72		&	-0.68		&1.35	&2.13\\
97072	&3.22	&0.55	&	&0.57	&0.90	&4.16		&	-0.20		&0.23	&3.84\\
97076	&4.59	&$\geq$0.88	&&0.56	&0.36	&1.63		&	0.20			&0.02	&15.94\\
97078	&5.44	&$\geq$0.96	&&0.61	&$\leq$0.06	&$\leq$0.30	&	$\geq$0.99	&-&-\\
97079	&2.34	&0.25	&	&0.25	&1.07	&2.65		&	-0.64		&1.06	&1.01\\
97082	&3.51	&$\geq$0.77	&&0.86	&$\leq$0.05	&$\leq$0.34	&	$\geq$1.20	&-&-\\
97091	&2.94	&-0.23	&	&0.66	&1.40	&7.18		&	-0.33		&0.59	&2.39\\
97092(S)	&2.49	&0.62	&	&0.32	&1.11	&3.37		&	-0.54		&0.21	&5.22\\
97093	&3.59	&$\geq$0.78	&&0.28	&0.50	&1.35		&	-0.26		&0.15	&3.43\\
97102(N)	&2.81	&0.41	&	&0.56	&0.52	&2.35		&	0.04		&0.06	&8.77\\
97111(S)	&0.96	&$\geq$0.21	&&0.29	&$\leq$0.12	&$\leq$0.35	&	$\geq$0.38	&-&-\\
97114	&1.32	&$\geq$0.34	&&0.24	&0.80	&1.94		&	-0.52		&0.28	&2.86\\
97120	&3.57	&$\geq$0.78	&&0.84	&1.41	&8.66		&	-0.23			&0.27	&5.25\\
97121	&3.19	&0.38	&	&0.8		&0.27	&1.58		&	0.48			&0.20	&1.30\\
97122	&7.36	&0.47	&	&0.48	&1.65	&6.67		&	-0.54		&1.08	&1.52\\
97125	&2.09	&$\geq$-0.14	&&0.53	&0.65	&2.85		&	-0.09		&0.43	&1.54\\
97129(W)	&10.7	&-0.07	&	&1.21	&0.98	&4.05		&	0.09			&1.58	&0.61\\
97073*	&2.39	&0.02	&	&0.24	&0.94	&2.3	0		&	-0.59		&1.08	&0.87\\
97087*	&12.89	&0.39	&	&0.67	&2.03	&11.7		&	-0.48		&3.67	&0.55\\
\hline

\end{tabular}
\end{minipage}
\end{table*}

\begin{table}
\centering
\begin{minipage}{140mm}
\caption{Spiral deficiency class parameters}
\label{gasdef}
\begin{tabular}[b]{@{}llrlrl@{}}
\hline
Class & \\
 
\hline\\
non--deficient &-0.3$ \leq$ \dhi\   $\leq$0.3&&  -0.4$ \leq$ \hd\   $\leq$0.4 & \\
gas deficient &\hspace{7mm} \dhi\   $\geq$0.3&&\hspace{7mm}  \hd\   $\geq$0.4 &\\
\hi\ deficient	&\hspace{7mm} \dhi\   $\geq$0.3&&  -0.4$ \leq$ \hd\   $\leq$0.4 &\\
\hdos\ deficient&-0.3$ \leq$ \dhi\   $\leq$0.3&& \hspace{7mm}  \hd\   $\geq$0.4 &\\
\hline
\end{tabular}
\end{minipage}
\end{table}

\begin{table}
\centering
\begin{minipage}{140mm}
\caption{\textcolor{black}{ISM deficiencies and colour by evolutionary state}}
\label{ttt}
\begin{tabular}{@{}ccclll@{}}
\hline
State\footnote{\textcolor{black}{For the parameters defining each evolutionary state see Table \ref{estates}}} &Plot&number&\dhi &\textit{g-i} colour&\hd\footnote{Deficiencies are from our IRAM observations except for \\ \cg129W where \hd\ was derived from the  Kitt Peak \\ flux from \cite{bosel97}. } \\
&symbol  &galaxies&median &median&mean  \\
\hline 

A& $\ast$&6&0.14 &0.59&$-0.39 \pm 0.27$\\ 
B&  \hspace{1.25mm}\raisebox{.6ex}{\circle{4.0}}&7&0.47&0.99&$-0.46 \pm 0.19$\\ 
C &  \textbullet&4&0.30&1.20 &\ \ $0.25 \pm 0.22$\\ 
D &  \framebox(4,4){×}  & 6&0.79&1.15&$\ \ 0.38 \pm 0.68$ \\ 
\hline
\end{tabular}
\end{minipage}
\end{table}

\begin{figure*}
\begin{minipage}{\linewidth}
\centering
\includegraphics[scale=0.65] {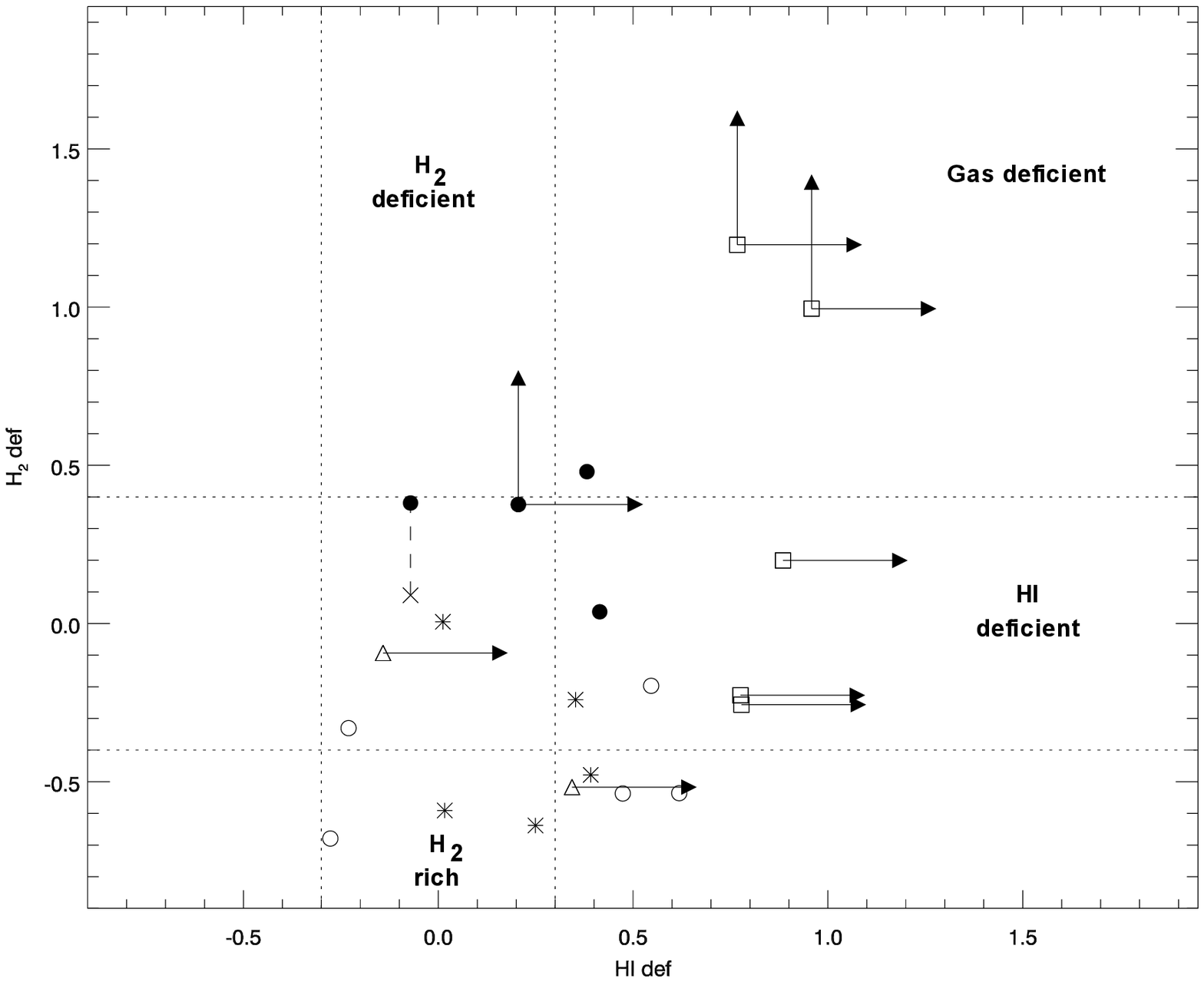}
\caption{\textbf{ \hi\ and \hdos\ deficiencies } The evolutionary state of the spirals as per Paper I are shown with the following symbols, State A  ($*$), including \cg073 and \cg087 from Boselli et al (1994), State B (o), State C  (\textbullet), State D ($\Box$) and BIG ($\bigtriangleup$). Arrows indicate lower limits for non--detections. For \cg129W the \hdos\ deficiency from \citep{bosel97} using the Kitt Peak 12--m telescope is indicated with a $\times$ and linked to our IRAM 30--m value  with the dashed line. The dotted lines give estimated upper and lower limits of natural variations in \hdos\ (0.4 dex) and \defhi\ (0.3 dex).  These lines divide the plot into a central box of natural variation and other regions labelled according to their gas content deficiency or excess. \hi\ data are taken from Paper I.}
\label{defcomp}

\end{minipage}
\end{figure*}

Notable results regarding their \hdos\ content, including significant differences with previous results, follow:

\begin{itemize}
\item 
\cg082 was not detected in CO at either 2.6 mm or 1.3 mm giving an upper \hdos\ mass limit of 4 $\times 10^7$ \msolar.  However \cite{bosel97} reported an indicative \hdos\ mass  of 4.8 $\times 10^8$ \msolar\ and a velocity of 6057\,\km\ using the NRAO Kitt Peak  12--m. We have confirmed our pointing position with NED and GOLDMine and our observations were centred at the optical velocity of 6100\,\km.  To explain the discrepancy in terms of a pointing error would require a pointing error 5--6 times greater than our estimated  pointing uncertainty of 2 arcsecs. While we believe our results to be correct, an independent observation is required to settle the issue.


\item
The \hdos\ mass derived from the  \hoz\ central pointing  \cg129W Kitt Peak 12--m  is 0.98 x 10$^9$ nearly double that derived from the \hoz\  central pointing with the IRAM 30--m.  This appears to be a clear indication that the IRAM beam resolved the \hoz\ emission and a substantial amount of \hdos\ lies beyond the IRAM beam (see also Appendices A and C).    
  
 \end{itemize}

Overall we find a range of \hdos\ deficiencies, with a stronger correlation between \textcolor{black}{\hdos\ deficiency and } evolutionary type than  type \hi\ deficiency.

\subsection{Star formation}
\label{sfrs}

For the sample spirals detected in \halpha, Table \ref{calc} shows their current SFR(\halpha) using the method from \cite{bosel02}. Corrections (dependent on morphological type) were made for [NII] contamination and internal extinction as per \cite{bosel01}. Where FIR fluxes were available we also calculated the  SFR(FIR)  using the method from \cite{kew02}; they are in reasonable agreement with the SFR(\halpha). The only exception was \cg111(S), which had a SFR(FIR) of 2.0 \msolaryr. We did not include this value in the table because it was inconsistent with the \halpha, CO and \hi\ non-detections; neither \cg111(S) nor its companion  are listed as separate entries in the IRAS PSC catalogue and it is likely the single entry suffers from confusion. The fluxes used to determine SFR(\halpha) and SFR(FIR) are from GOLDMine.

Table \ref{calc} also lists the \hdos\ depletion time, defined as the current molecular mass over the current SFR, i.e., no correction is applied for replenishment. These depletion times are in line with what has been found for galaxies observed as part of The \hi\ Nearby Galaxies Survey \citep{leroy08}. We see that the current SFR would consume the current \hdos\ reservoirs of the evolutionary state A spirals on time scales of 10$^9$ yr or less, confirming their classification as starburst galaxies, under the definition of a starburst as a galaxy with a molecular depletion timescale smaller than the age of the Universe. Those grouped in evolutionary state D have no or little detected current star formation and therefore, even though their molecular gas masses are low, show depletion time scales of several Gyrs. As expected the EW(\halpha+[NII]) from Table \ref{pramss} plus values from GOLDMine for \cg073 and \cg087 show a trend in the sense that less evolved spirals display larger EW(\halpha+[NII]) values than the more evolved ones (Spearman coefficient $r_s$ =0.83).

\subsection{Constraints on the size and shape of the molecular discs}
\label{maps}

Strong tidal interactions are known to perturb molecular discs \citep{iono05}. However, it is not yet clear observationally how these discs are affected by the high levels of ram pressure that they experience in galaxy clusters. With this is mind, we investigate whether the CO distribution in our galaxies is similar to that seen in unperturbed nearby spirals. Since we lack complete maps of the CO emission, we can only check whether our observations are consistent with this hypothesis. Following the studies of local galaxies by, e.g., \cite{young95} and \cite{leroy09}, we model each molecular disc as an axisymmetric exponential distribution	with	a	scale	length	in	the	range	$\sim$	0.10\textcolor{black}{$r_{25}$} -- 0.35\textcolor{black}{$r_{25}$}\footnote{\cite{leroy09} find scale lengths in the range $\sim$ 0.10\textcolor{black}{$r_{25}$} -- 0.32\textcolor{black}{$r_{25}$} with an average value of 0.2\textcolor{black}{$r_{25}$.}}. Its centre, inclination and orientation are as in the optical disc. The model disc is then convolved with the beam of the telescope to simulate the observations \textcolor{black}{ allowing us to make the following comparison with the observed profiles.}

We observed eight galaxies at multiple positions and in Appendix A we comment on the observed molecular gas distributions in each of them. In each of these objects we fit a Gaussian to the \hoz\ and \hto\ intensities measured along every direction that contains three or more aligned pointings. Each fit yields the FWHM and the offset of the peak of the Gaussian profile along the scanning direction. The exponential disc convolved with a Gaussian beam can be well described as a Gaussian disc at the galactocentric radii that we sample, provided that the beam size is larger than the disc scale length. This is not necessarily the case for \hto\ observations. Therefore, we fit the profiles of both transitions but only the \hoz\ data are compared with the values expected from the simulated discs.

Table \ref{gauss} summarizes the results of the fits. In Figure \ref{gffwhm}, we compare the FWHMs from the fits with the values expected	assuming	scale	lengths	in	the	range	0.10\textcolor{black}{$r_{25}$ --0.35$r_{25}$}. \textcolor{black}{For the three outliers in Figure \ref{gffwhm}, \cg062, \cg079 and \cg102(N) the 3 $\sigma$ value for the fits to the size of their CO disks are  $\lesssim$  $0.05r_{25}$, $\gtrsim$  $r_{25}$ and $\gtrsim$  $0.5r_{25}$ respectively. These values should be treated with caution as the assumption of a regular, exponential disk is most likely invalid for these spirals. Setting} aside the cases of \cg079 and \cg102(N), discussed below, the observations are in good agreement with the outcome of the models. On the one hand, the peak offsets are of the order of the pointing uncertainties or less and typically compatible with zero within errors. Therefore we find no significant evidence of asymmetry in the CO distributions. On the other hand, all the FWHMs of the fits, except for \cg062, are compatible with scale lengths within the range observed in nearby spirals. In the case of \cg062, the fitted Gaussian is roughly as wide as the 22 arcsec beam, which suggests \textcolor{black}{either the CO disk is very small or there is a significant, very compact component (e.g., a nuclear disk) in addition to the disk}.

We now review the results for \cg079 and \cg102(N). For these galaxies, the fits yield very uncertain and large offsets, as well as unexpectedly large FWHMs. Essentially, the Gaussian approach fails in both cases. In \cg079, the \hoz\ observations are compatible, within errors, with a nearly flat radial profile that hints at an off-centre peak at $\sim$ 17 arcsec. This could be explained by a large, axisymmetric disc or, more simply, by a strong asymmetry NW from the optical centre. In \cg102(N), the measured CO intensity increases monotonically from SE to NW across the centre. Possible scenarios are a misalignment between the optical and molecular discs or an asymmetry along the scanning direction.

In summary, thirteen of the galaxies in our sample are subject to at least one of the two tests described above. In three of them we find convincing evidence that the shape and/or size of their molecular discs differs from that of nearby unperturbed spirals. In \cg62, the CO emission is much more compact than expected. In \cg079 and \cg102(N), the CO distributions seem to be significantly shifted from the optical centres.

\begin{table*} 
\centering
\begin{minipage}{90mm}
\caption{Gaussian fits for spirals with multiple CO detections}
\label{gauss}
\begin{tabular}[h]{@{}llrrrrr@{}}
\hline
 &Orientation& &CO(1--0)&  &&CO(2--1) \\
 \cline{3-4} \cline{6-7}
CGCG &of pointings &mean\footnote{Offset in arcsec from the optical centre of the galaxy from fitting a single Gaussian to the detections along the axis in the second column. The direction of the offsets is positive to the North except for \cg092(S) where positive is to the East.  }&FWHM\footnote{FWHM of a single Gaussian fit to detections along the axis in the second column. }& & mean$^a$ &FWHM$^b$\\
& [PA\degree]&[arcsec]&[arcsec]&& [arcsec]&[arcsec] \\
\hline\\

97062&45&-0.6 $\pm$\,1.2&18.7$\pm$\,1.4&&\,0.6$\pm$ 1.2&\,18.8$\pm$1.2\\
97068&0&-1.7 $ \pm$ \,0.2&24.5$\pm$\,0.6&&-2.2$\pm$0.2&16.6$\pm$0.6\\
97072& 127&3.4 $ \pm$ \,0.9& 31.6$\pm$\,1.5&&1.2$\pm$0.9&19.3$\pm$1.2\\
97079\footnote{  \hoz\ fit to our data plus the 0, 0 and  -10, 10 data points from \cite{bosel94}.  }&315&15.6 $\pm$ 8.5&62.0$\pm$ 5.9&&5.2$\pm$3.7&34.1$\pm$3.0\\
97092(S)&0& -0.1 $\pm$ \,0.5&26.4$\pm$\,1.3&&0.3$\pm$0.5&16.5$\pm$0.9\\
97092(S)&90& -1.9 $\pm$ \,0.7&25.1$\pm$\,1.4&&-&-\\
97102(N)&139& 11.1 $\pm$11.2& 56.9$\pm$5.2&&-&-\\
97120&218&-1.7  $\pm$ \,1.0&40.0$\pm$\,1.8&&0.9$\pm$0.6&20.3$\pm$1.0\\
97129(W)&192&1.1 $\pm$ \,1.4& 32.2$\pm$\,2.1&&-&-\\
\hline
\end{tabular}
\end{minipage}
\end{table*}

\begin{figure}
\begin{center}
\includegraphics[ angle=0,scale=0.43] {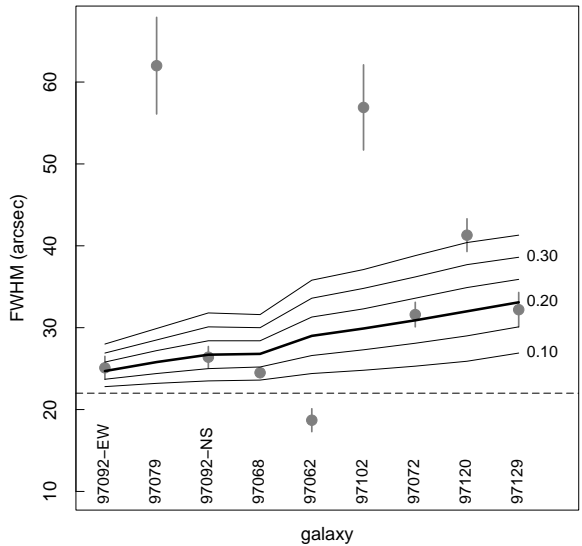}
\vspace{1cm}
\caption{FWHM of the Gaussian profiles fitted to the multi- pointing \hoz\ observations. The galaxies are ordered from left to right in increasing order of the size of the optical disk along the scanning direction (assuming elliptical discs; note that the North-South and East-West scans across \cg092(S) have different projected sizes). The grey dots and error bars indicate the results from the fits and their $\pm$1$\sigma$ uncertainty. The solid lines join the FWHM values expected from the models assuming CO scale lengths from 0.10 to 0.35 times \textcolor{black}{$r_{25}$} in steps of 0.05\textcolor{black}{$r_{25}$}, as indicated on the right-hand side. The dashed line corresponds to the 30m telescope resolution (22 arcsec).}
\label{gffwhm}
\end{center}
\end{figure}

\section{Concluding remarks}
In summary:
\begin{itemize}
\item
The mean \hdos\ deficiencies of the spirals with more evolved evolutionary states C and D (red and moderately or severely \hi\ deficient) are significantly larger than for the less evolved the state A and B spirals (blue and moderately or not significantly \hi\ deficient). Two state D spirals (\cg078 and \cg082) were found to be exceptionally gas poor (i.e., highly deficient in both \hi\ and \hdos).
\item
The differences between \hdos\ deficiency in the two evolutionary types, A\&B and C\&D, (based on SDSS \textit{g--i }colour and	\dhi)	and	lack	of	correlation with \dhi\ argues	for a connection between \hdos\ deficiency and a process which is more closely associated with colour evolution than ram pressure stripping, thought to be responsible for the \hi\ deficiencies. At the current SFRs all the state A spirals would deplete their \hdos\ reservoirs within the order of 10$^9$ yr, implying that for these spirals at least, star formation has the potential to be a significant cause of future molecular gas depletion.

 \item
Our central pointing results, the multiple pointings observations and the KP--vs--30m comparisons in particular, are consistent with most of the observed galaxies having exponential CO disks of scale lengths similar to those in nearby objects. The clearest exceptions are \cg079 and \cg102(N), which have significant offsets between their CO and optical intensity maxima of 6 and 4.5 kpc, respectively. Two possible further exceptions are \cg062: with multiple \hoz\ pointings suggesting a more compact CO disk than expected, although the SNR at the off--center positions have a low SNR, and \cg121: the Kitt Peak/ IRAM 30--m flux ratio is $\sim$1.8 times the expected value which could be due to either instrumental errors or a more extended CO disk.
\item
For \cg079 the perturbed NIR stellar and molecular morphologies and short \hdos\ depletion time scale suggest a tidal interaction. But the narrow 75 kpc \halpha\ and radio continuum tails, and the \hi\ offset favour the view that the spiral is subject to significant ram pressure stripping. The indication from the single dish observations of an \hdos\ intensity offset of $\sim$ 6 kpc in the tail direction does not readily accord with the existing pure hydrodynamic interaction models, so a hydrodynamic plus tidal interaction cannot be ruled out as the cause of the offset.
\end{itemize}

Despite the uncertainties associated with our single dish observations we find that spirals in A\,1367 have wide a range of \hdos\ deficiencies that are correlated with the evolutionary states of the spirals. The cases of \cg079 and \cg129W \textcolor{black}{deviate from} the commonly assumed exponential distribution of molecular gas in spirals. Both of these results point to the importance of high spatial resolution mapping of molecular gas to advance our understanding of the processes by which the cluster environment impacts the evolution of cluster spirals. Specifically, spatially resolved CO studies might be a means to discriminate between hydrodynamic and tidal interaction mechanisms in the realm of moderate to low ram pressure.

\label{conclusion}

\section*{Acknowledgments}

HBA acknowledges support for this project via CONACyT grant No.\ 50794.

This research has made use of the NASA/IPAC Extragalactic Database (NED) which is operated by the Jet Propulsion Laboratory, California Institute of Technology, under contract with the National Aeronautics and Space Administration.

This research has made use of the Sloan Digital Sky Survey (SDSS). Funding for the SDSS and SDSS-II has been provided by the Alfred P. Sloan Foundation, the Participating Institutions, the National Science Foundation, the U.S. Department of Energy, the National Aeronautics and Space Administration, the Japanese Monbukagakusho, the Max Planck Society, and the Higher Education Funding Council for England. The SDSS Web Site is http://www.sdss.org/.

The research leading to these results has received funding from the European Community's Seventh Framework Programme (/FP7/2007-2013/) under grant agreement No 229517.

\bibliographystyle{mn2e}
\bibliography{cluster}

\begin{thebibliography}{}

\bibitem[\protect\citeauthoryear{{Bekki}}{{Bekki}}{1999}]{bekki99}
{Bekki} K.,  1999, ApJL, 510, L15

\bibitem[\protect\citeauthoryear{{Bekki}, {Couch} \& {Shioya}}{{Bekki}
  et~al.}{2002}]{bekki02}
{Bekki} K.,  {Couch} W.~J.,    {Shioya} Y.,  2002, ApJ, 577, 651

\bibitem[\protect\citeauthoryear{{Boselli}, {Boissier}, {Cortese}, {Gil de
  Paz}, {Seibert}, {Madore}, {Buat} \& {Martin}}{{Boselli}
  et~al.}{2006}]{bosel06b}
{Boselli} A.,  {Boissier} S.,  {Cortese} L.,  {Gil de Paz} A.,  {Seibert} M.,
  {Madore} B.~F.,  {Buat} V.,    {Martin} D.~C.,  2006, ApJ, 651, 811

\bibitem[\protect\citeauthoryear{{Boselli} \& {Gavazzi}}{{Boselli} \&
  {Gavazzi}}{2006}]{bosel06a}
{Boselli} A.,  {Gavazzi} G.,  2006, PASP, 118, 517

\bibitem[\protect\citeauthoryear{{Boselli}, {Gavazzi}, {Combes} \&
  {Lequeux}}{{Boselli} et~al.}{1994}]{bosel94}
{Boselli} A.,  {Gavazzi} G.,  {Combes} F.,    {Lequeux} J.,  1994, A\&A, 285,
  69

\bibitem[\protect\citeauthoryear{{Boselli}, {Gavazzi}, {Donas} \&
  {Scodeggio}}{{Boselli} et~al.}{2001}]{bosel01}
{Boselli} A.,  {Gavazzi} G.,  {Donas} J.,    {Scodeggio} M.,  2001, AJ, 121,
  753

\bibitem[\protect\citeauthoryear{{Boselli}, {Gavazzi}, {Lequeux} \&
  {Buat}}{{Boselli} et~al.}{1997}]{bosel97}
{Boselli} A.,  {Gavazzi} G.,  {Lequeux} J.,    {Buat} V.,  1997, A\&A, 327, 522

\bibitem[\protect\citeauthoryear{{Boselli}, {Lequeux} \& {Gavazzi}}{{Boselli}
  et~al.}{2002}]{bosel02}
{Boselli} A.,  {Lequeux} J.,    {Gavazzi} G.,  2002, A\&A, 384, 33

\bibitem[\protect\citeauthoryear{{Broeils} \& {Rhee}}{{Broeils} \&
  {Rhee}}{1997}]{broeils97}
{Broeils} A.~H.,  {Rhee} M.,  1997, A\&A, 324, 877

\bibitem[\protect\citeauthoryear{{Byrd} \& {Valtonen}}{{Byrd} \&
  {Valtonen}}{1990}]{byrd90}
{Byrd} G.,  {Valtonen} M.,  1990, ApJ, 350, 89

\bibitem[\protect\citeauthoryear{{Casoli}, {Boisse}, {Combes} \&
  {Dupraz}}{{Casoli} et~al.}{1991}]{casoli91}
{Casoli} F.,  {Boisse} P.,  {Combes} F.,    {Dupraz} C.,  1991, A\&A, 249, 359

\bibitem[\protect\citeauthoryear{{Cayatte}, {Kotanyi}, {Balkowski} \& {van
  Gorkom}}{{Cayatte} et~al.}{1994}]{caya94}
{Cayatte} V.,  {Kotanyi} C.,  {Balkowski} C.,    {van Gorkom} J.~H.,  1994, AJ,
  107, 1003

\bibitem[\protect\citeauthoryear{{Combes}, {Baker}, {Schinnerer},
  {Garc{\'{\i}}a-Burillo}, {Hunt}, {Boone}, {Eckart}, {Neri} \&
  {Tacconi}}{{Combes} et~al.}{2009}]{combes09}
{Combes} F.,  {Baker} A.~J.,  {Schinnerer} E.,  {Garc{\'{\i}}a-Burillo} S.,
  {Hunt} L.~K.,  {Boone} F.,  {Eckart} A.,  {Neri} R.,    {Tacconi} L.~J.,
  2009, A\&A, 503, 73

\bibitem[\protect\citeauthoryear{{Cort{\'e}s}, {Kenney} \&
  {Hardy}}{{Cort{\'e}s} et~al.}{2006}]{cortes06}
{Cort{\'e}s} J.~R.,  {Kenney} J.~D.~P.,    {Hardy} E.,  2006, AJ, 131, 747

\bibitem[\protect\citeauthoryear{{Cortese}, {Boselli}, {Gavazzi},
  {Iglesias-Paramo}, {Madore}, {Barlow} \& {Bianchi}}{{Cortese}
  et~al.}{2005}]{cort05}
{Cortese} L.,  {Boselli} A.,  {Gavazzi} G.,  {Iglesias-Paramo} J.,  {Madore}
  B.~F.,  {Barlow} T.,    {Bianchi} L.,  2005, ApJL, 623, L17

\bibitem[\protect\citeauthoryear{{Cortese}, {Gavazzi}, {Boselli}, {Franzetti},
  {Kennicutt}, {O'Neil} \& {Sakai}}{{Cortese} et~al.}{2006}]{cort06}
{Cortese} L.,  {Gavazzi} G.,  {Boselli} A.,  {Franzetti} P.,  {Kennicutt}
  R.~C.,  {O'Neil} K.,    {Sakai} S.,  2006, A\&A, 453, 847

\bibitem[\protect\citeauthoryear{{Cortese}, {Minchin}, {Auld}, {Davies},
  {Catinella}, {Momjian}, {Rosenberg}, {Taylor}, {Gavazzi}, {O'Neil}, {Baes},
  {Boselli}, {Bothun}, {Koribalski}, {Schneider} \& {van Driel}}{{Cortese}
  et~al.}{2008}]{cort08}
{Cortese} L.,  {Minchin} R.~F.,  {Auld} R.~R.,  {Davies} J.~I.,  {Catinella}
  B.,  {Momjian} E.,  {Rosenberg} J.~L.,  {Taylor} R.,  {Gavazzi} G.,  {O'Neil}
  K.,  {Baes} M.,  {Boselli} A.,  {Bothun} G.,  {Koribalski} B.,  {Schneider}
  S.,    {van Driel} W.,  2008, MNRAS, 383, 1519

\bibitem[\protect\citeauthoryear{{Dickey} \& {Kazes}}{{Dickey} \&
  {Kazes}}{1992}]{dick92}
{Dickey} J.~M.,  {Kazes} I.,  1992, ApJ, 393, 530

\bibitem[\protect\citeauthoryear{{Fujita} \& {Goto}}{{Fujita} \&
  {Goto}}{2004}]{fuji04}
{Fujita} Y.,  {Goto} T.,  2004, PASJ, 56, 621

\bibitem[\protect\citeauthoryear{{Fumagalli}, {Krumholz}, {Prochaska},
  {Gavazzi} \& {Boselli}}{{Fumagalli} et~al.}{2009}]{fuma09}
{Fumagalli} M.,  {Krumholz} M.~R.,  {Prochaska} J.~X.,  {Gavazzi} G.,
  {Boselli} A.,  2009, ApJ, 697, 1811

\bibitem[\protect\citeauthoryear{{Gavazzi}, {Boselli}, {Donati}, {Franzetti} \&
  {Scodeggio}}{{Gavazzi} et~al.}{2003}]{gava03b}
{Gavazzi} G.,  {Boselli} A.,  {Donati} A.,  {Franzetti} P.,    {Scodeggio} M.,
  2003, A\&A, 400, 451

\bibitem[\protect\citeauthoryear{{Gavazzi}, {Boselli}, {Mayer},
  {Iglesias-Paramo}, {V{\'{\i}}lchez} \& {Carrasco}}{{Gavazzi}
  et~al.}{2001}]{gava01b}
{Gavazzi} G.,  {Boselli} A.,  {Mayer} L.,  {Iglesias-Paramo} J.,
  {V{\'{\i}}lchez} J.~M.,    {Carrasco} L.,  2001, ApJL, 563, L23

\bibitem[\protect\citeauthoryear{{Gavazzi}, {Boselli}, {Pedotti}, {Gallazzi} \&
  {Carrasco}}{{Gavazzi} et~al.}{2002}]{gava02}
{Gavazzi} G.,  {Boselli} A.,  {Pedotti} P.,  {Gallazzi} A.,    {Carrasco} L.,
  2002, A\&A, 396, 449

\bibitem[\protect\citeauthoryear{{Gavazzi}, {Franzetti}, {Scodeggio}, {Boselli}
  \& {Pierini}}{{Gavazzi} et~al.}{2000}]{gava00}
{Gavazzi} G.,  {Franzetti} P.,  {Scodeggio} M.,  {Boselli} A.,    {Pierini} D.,
   2000, A\&A, 361, 863

\bibitem[\protect\citeauthoryear{{Gavazzi} \& {Jaffe}}{{Gavazzi} \&
  {Jaffe}}{1987}]{gava87}
{Gavazzi} G.,  {Jaffe} W.,  1987, A\&A, 186, L1

\bibitem[\protect\citeauthoryear{{Gavazzi} \& {Scodeggio}}{{Gavazzi} \&
  {Scodeggio}}{1996}]{gava96}
{Gavazzi} G.,  {Scodeggio} M.,  1996, A\&A, 312, L29

\bibitem[\protect\citeauthoryear{{Gunn} \& {Gott}}{{Gunn} \&
  {Gott}}{1972}]{gun72}
{Gunn} J.~E.,  {Gott} J.~R.~I.,  1972, ApJ, 176, 1

\bibitem[\protect\citeauthoryear{{Haynes} \& {Giovanelli}}{{Haynes} \&
  {Giovanelli}}{1984}]{hayn84}
{Haynes} M.~P.,  {Giovanelli} R.,  1984, AJ, 89, 758

\bibitem[\protect\citeauthoryear{{Hota} \& {Saikia}}{{Hota} \&
  {Saikia}}{2007}]{hota07}
{Hota} A.,  {Saikia} D.~J.,  2007, Bulletin of the Astronomical Society of
  India, 35, 121

\bibitem[\protect\citeauthoryear{{Iglesias-P{\'a}ramo}, {Boselli}, {Cortese},
  {V{\'{\i}}lchez} \& {Gavazzi}}{{Iglesias-P{\'a}ramo} et~al.}{2002}]{ipara02}
{Iglesias-P{\'a}ramo} J.,  {Boselli} A.,  {Cortese} L.,  {V{\'{\i}}lchez}
  J.~M.,    {Gavazzi} G.,  2002, A\&A, 384, 383

\bibitem[\protect\citeauthoryear{{Iono}, {Yun} \& {Ho}}{{Iono}
  et~al.}{2005}]{iono05}
{Iono} D.,  {Yun} M.~S.,    {Ho} P.~T.~P.,  2005, ApJS, 158, 1

\bibitem[\protect\citeauthoryear{{Kapferer}, {Kronberger}, {Ferrari}, {Riser}
  \& {Schindler}}{{Kapferer} et~al.}{2008}]{kapf08}
{Kapferer} W.,  {Kronberger} T.,  {Ferrari} C.,  {Riser} T.,    {Schindler} S.,
   2008, MNRAS, 389, 1405

\bibitem[\protect\citeauthoryear{{Kapferer}, {Sluka}, {Schindler}, {Ferrari} \&
  {Ziegler}}{{Kapferer} et~al.}{2009}]{kapf09}
{Kapferer} W.,  {Sluka} C.,  {Schindler} S.,  {Ferrari} C.,    {Ziegler} B.,
  2009, A\&A, 499, 87

\bibitem[\protect\citeauthoryear{{Kaufman}, {Sheth}, {Struck}, {Elmegreen},
  {Thomasson}, {Elmegreen} \& {Brinks}}{{Kaufman} et~al.}{2002}]{kauf02}
{Kaufman} M.,  {Sheth} K.,  {Struck} C.,  {Elmegreen} B.~G.,  {Thomasson} M.,
  {Elmegreen} D.~M.,    {Brinks} E.,  2002, AJ, 123, 702

\bibitem[\protect\citeauthoryear{{Kennicutt} Jr.}{{Kennicutt}}{1983}]{kenni83}
{Kennicutt} Jr. R.~C.,  1983, AJ, 88, 483

\bibitem[\protect\citeauthoryear{{Kewley}, {Geller}, {Jansen} \&
  {Dopita}}{{Kewley} et~al.}{2002}]{kew02}
{Kewley} L.~J.,  {Geller} M.~J.,  {Jansen} R.~A.,    {Dopita} M.~A.,  2002, AJ,
  124, 3135

\bibitem[\protect\citeauthoryear{{Komugi}, {Kohno}, {Tosaki}, {Nakanishi},
  {Onodera}, {Egusa} \& {Sofue}}{{Komugi} et~al.}{2007}]{komugi07}
{Komugi} S.,  {Kohno} K.,  {Tosaki} T.,  {Nakanishi} H.,  {Onodera} S.,
  {Egusa} F.,    {Sofue} Y.,  2007, PASJ, 59, 55

\bibitem[\protect\citeauthoryear{{Koopmann} \& {Kenney}}{{Koopmann} \&
  {Kenney}}{2004}]{koop04}
{Koopmann} R.~A.,  {Kenney} J.~D.~P.,  2004, ApJ, 613, 851

\bibitem[\protect\citeauthoryear{{Kronberger}, {Kapferer}, {Ferrari},
  {Unterguggenberger} \& {Schindler}}{{Kronberger} et~al.}{2008}]{kronb08}
{Kronberger} T.,  {Kapferer} W.,  {Ferrari} C.,  {Unterguggenberger} S.,
  {Schindler} S.,  2008, A\&A, 481, 337

\bibitem[\protect\citeauthoryear{{Krumholz}, {McKee} \& {Tumlinson}}{{Krumholz}
  et~al.}{2008}]{kurmh08}
{Krumholz} M.~R.,  {McKee} C.~F.,    {Tumlinson} J.,  2008, ApJ, 689, 865

\bibitem[\protect\citeauthoryear{{Krumholz}, {McKee} \& {Tumlinson}}{{Krumholz}
  et~al.}{2009}]{krumh09}
{Krumholz} M.~R.,  {McKee} C.~F.,    {Tumlinson} J.,  2009, ApJ, 693, 216

\bibitem[\protect\citeauthoryear{{Larson}, {Tinsley} \& {Caldwell}}{{Larson}
  et~al.}{1980}]{larson80}
{Larson} R.~B.,  {Tinsley} B.~M.,    {Caldwell} C.~N.,  1980, ApJ, 237, 692

\bibitem[\protect\citeauthoryear{{Lavezzi}, {Dickey}, {Casoli} \&
  {Kaz{\`e}s}}{{Lavezzi} et~al.}{1999}]{lave99}
{Lavezzi} T.~E.,  {Dickey} J.~M.,  {Casoli} F.,    {Kaz{\`e}s} I.,  1999, AJ,
  117, 1995

\bibitem[\protect\citeauthoryear{{Leroy}, {Walter}, {Bigiel}, {Usero}, {Weiss},
  {Brinks}, {de Blok}, {Kennicutt}, {Schuster}, {Kramer}, {Wiesemeyer} \&
  {Roussel}}{{Leroy} et~al.}{2009}]{leroy09}
{Leroy} A.~K.,  {Walter} F.,  {Bigiel} F.,  {Usero} A.,  {Weiss} A.,  {Brinks}
  E.,  {de Blok} W.~J.~G.,  {Kennicutt} R.~C.,  {Schuster} K.-F.,  {Kramer} C.,
   {Wiesemeyer} H.~W.,    {Roussel} H.,  2009, AJ, 137, 4670

\bibitem[\protect\citeauthoryear{{Leroy}, {Walter}, {Brinks}, {Bigiel}, {de
  Blok}, {Madore} \& {Thornley}}{{Leroy} et~al.}{2008}]{leroy08}
{Leroy} A.~K.,  {Walter} F.,  {Brinks} E.,  {Bigiel} F.,  {de Blok} W.~J.~G.,
  {Madore} B.,    {Thornley} M.~D.,  2008, AJ, 136, 2782

\bibitem[\protect\citeauthoryear{{Lisenfeld}, {Espada}, {Verdes-Montenegro},
  {Kuno}, {Leon}, {Sabater}, {Sato}, {Sulentic}, {Verley} \& {Yun}}{{Lisenfeld}
  et~al.}{2011}]{lisenfeld11}
{Lisenfeld} U.,  {Espada} D.,  {Verdes-Montenegro} L.,  {Kuno} N.,  {Leon} S.,
  {Sabater} J.,  {Sato} N.,  {Sulentic} J.,  {Verley} S.,    {Yun} M.~S.,
  2011, A\&A, 534, A102

\bibitem[\protect\citeauthoryear{{Lucero}, {Young} \& {van Gorkom}}{{Lucero}
  et~al.}{2005}]{lucero05}
{Lucero} D.~M.,  {Young} L.~M.,    {van Gorkom} J.~H.,  2005, AJ, 129, 647

\bibitem[\protect\citeauthoryear{{Martig} \& {Bournaud}}{{Martig} \&
  {Bournaud}}{2008}]{mart08}
{Martig} M.,  {Bournaud} F.,  2008, MNRAS, 385, L38

\bibitem[\protect\citeauthoryear{{Martinez-Badenes}, {Lisenfeld}, {Espada},
  {Verdes-Montenegro}, {Garc{\'{\i}}a-Burillo}, {Leon}, {Sulentic} \&
  {Yun}}{{Martinez-Badenes} et~al.}{2012}]{badenes12}
{Martinez-Badenes} V.,  {Lisenfeld} U.,  {Espada} D.,  {Verdes-Montenegro} L.,
  {Garc{\'{\i}}a-Burillo} S.,  {Leon} S.,  {Sulentic} J.,    {Yun} M.~S.,
  2012, A\&A, 540, A96

\bibitem[\protect\citeauthoryear{{Mauersberger}, {Guelin}, {Martin-Pintado},
  {Thum}, {Cernicharo}, {Hein} \& {Navarro}}{{Mauersberger}
  et~al.}{1989}]{mauersberger89}
{Mauersberger} R.,  {Guelin} M.,  {Martin-Pintado} J.,  {Thum} C.,
  {Cernicharo} J.,  {Hein} H.,    {Navarro} S.,  1989, 79, 217

\bibitem[\protect\citeauthoryear{{Moore}, {Katz}, {Lake}, {Dressler} \&
  {Oemler}}{{Moore} et~al.}{1996}]{moore96}
{Moore} B.,  {Katz} N.,  {Lake} G.,  {Dressler} A.,    {Oemler} A.,  1996,
  Nature, 379, 613

\bibitem[\protect\citeauthoryear{{Moore}, {Lake}, {Quinn} \& {Stadel}}{{Moore}
  et~al.}{1999}]{moore99}
{Moore} B.,  {Lake} G.,  {Quinn} T.,    {Stadel} J.,  1999, MNRAS, 304, 465

\bibitem[\protect\citeauthoryear{{Natarajan}, {Kneib} \& {Smail}}{{Natarajan}
  et~al.}{2002}]{nata02}
{Natarajan} P.,  {Kneib} J.-P.,    {Smail} I.,  2002, ApJL, 580, L11

\bibitem[\protect\citeauthoryear{{Nishiyama}, {Nakai} \& {Kuno}}{{Nishiyama}
  et~al.}{2001}]{nishiyama01}
{Nishiyama} K.,  {Nakai} N.,    {Kuno} N.,  2001, PASJ, 53, 757

\bibitem[\protect\citeauthoryear{{Nulsen}}{{Nulsen}}{1982}]{nuls82}
{Nulsen} P.~E.~J.,  1982, MNRAS, 198, 1007

\bibitem[\protect\citeauthoryear{{Polk}, {Knapp}, {Stark} \& {Wilson}}{{Polk}
  et~al.}{1988}]{Polk88}
{Polk} K.~S.,  {Knapp} G.~R.,  {Stark} A.~A.,    {Wilson} R.~W.,  1988, ApJ,
  332, 432

\bibitem[\protect\citeauthoryear{{Roediger} \& {Br{\"u}ggen}}{{Roediger} \&
  {Br{\"u}ggen}}{2007}]{roed07}
{Roediger} E.,  {Br{\"u}ggen} M.,  2007, MNRAS, p.~765

\bibitem[\protect\citeauthoryear{{Roediger} \& {Hensler}}{{Roediger} \&
  {Hensler}}{2005}]{roed05}
{Roediger} E.,  {Hensler} G.,  2005, A\&A, 433, 875

\bibitem[\protect\citeauthoryear{{Schneider}, {Thuan}, {Magri} \&
  {Wadiak}}{{Schneider} et~al.}{1990}]{schnei90}
{Schneider} S.~E.,  {Thuan} T.~X.,  {Magri} C.,    {Wadiak} J.~E.,  1990, ApJS,
  72, 245

\bibitem[\protect\citeauthoryear{{Scott}, {Bravo-Alfaro}, {Brinks}, {Caretta},
  {Cortese}, {Boselli}, {Hardcastle}, {Croston} \& {Plauchu}}{{Scott}
  et~al.}{2010}]{scott10}
{Scott} T.~C.,  {Bravo-Alfaro} H.,  {Brinks} E.,  {Caretta} C.~A.,  {Cortese}
  L.,  {Boselli} A.,  {Hardcastle} M.~J.,  {Croston} J.~H.,    {Plauchu} I.,
  2010, MNRAS, 403, 1175

\bibitem[\protect\citeauthoryear{{Sivanandam}, {Rieke} \& {Rieke}}{{Sivanandam}
  et~al.}{2010}]{sivanandam10}
{Sivanandam} S.,  {Rieke} M.~J.,    {Rieke} G.~H.,  2010, ApJ, 717, 147

\bibitem[\protect\citeauthoryear{{Spergel}, {Bean}, {Dor{\'e}}, {Nolta},
  {Bennett}, {Dunkley}, {Hinshaw}, {Jarosik}, {Komatsu}, {Page}, {Peiris},
  {Verde}, {Halpern}, {Hill}, {Kogut}, {Limon}, {Wollack} \&
  {Wright}}{{Spergel} et~al.}{2007}]{sperg07}
{Spergel} D.~N.,  {Bean} R.,  {Dor{\'e}} O.,  {Nolta} M.~R.,  {Bennett} C.~L.,
  {Dunkley} J.,  {Hinshaw} G.,  {Jarosik} N.,  {Komatsu} E.,  {Page} L.,
  {Peiris} H.~V.,  {Verde} L.,  {Halpern} M.,  {Hill} R.~S.,  {Kogut} A.,
  {Limon} M.,  {Wollack} E.,    {Wright} E.~L.,  2007, ApJS, 170, 377

\bibitem[\protect\citeauthoryear{{Strong}, {Bloemen}, {Dame}, {Grenier},
  {Hermsen}, {Lebrun}, {Nyman}, {Pollock} \& {Thaddeus}}{{Strong}
  et~al.}{1988}]{strong88}
{Strong} A.~W.,  {Bloemen} J.~B.~G.~M.,  {Dame} T.~M.,  {Grenier} I.~A.,
  {Hermsen} W.,  {Lebrun} F.,  {Nyman} L.-A.,  {Pollock} A.~M.~T.,
  {Thaddeus} P.,  1988, A\&A, 207, 1

\bibitem[\protect\citeauthoryear{{Sun}, {Donahue}, {Roediger}, {Nulsen},
  {Voit}, {Sarazin}, {Forman} \& {Jones}}{{Sun} et~al.}{2010}]{sun10}
{Sun} M.,  {Donahue} M.,  {Roediger} E.,  {Nulsen} P.~E.~J.,  {Voit} G.~M.,
  {Sarazin} C.,  {Forman} W.,    {Jones} C.,  2010, ApJ, 708, 946

\bibitem[\protect\citeauthoryear{{van Gorkom}}{{van Gorkom}}{2004}]{vgork04}
{van Gorkom} J.~H.,  2004, in {Mulchaey} J.~S.,  {Dressler} A.,   {Oemler} A.,
  eds, Clusters of Galaxies: Probes of Cosmological Structure and Galaxy
  Evolution {Interaction of Galaxies with the Intracluster Medium}.
p.~305

\bibitem[\protect\citeauthoryear{{Vollmer}, {Braine}, {Balkowski}, {Cayatte} \&
  {Duschl}}{{Vollmer} et~al.}{2001}]{voll01b}
{Vollmer} B.,  {Braine} J.,  {Balkowski} C.,  {Cayatte} V.,    {Duschl} W.~J.,
  2001, A\&A, 374, 824

\bibitem[\protect\citeauthoryear{{Vollmer}, {Braine}, {Pappalardo} \&
  {Hily-Blant}}{{Vollmer} et~al.}{2008}]{voll08}
{Vollmer} B.,  {Braine} J.,  {Pappalardo} C.,    {Hily-Blant} P.,  2008, A\&A,
  491, 455

\bibitem[\protect\citeauthoryear{{Young} \& {Knezek}}{{Young} \&
  {Knezek}}{1989}]{young89}
{Young} J.~S.,  {Knezek} P.~M.,  1989, ApJL, 347, L55

\bibitem[\protect\citeauthoryear{{Young}, {Xie}, {Tacconi}, {Knezek},
  {Viscuso}, {Tacconi-Garman}, {Scoville} \& {Schneider}}{{Young}
  et~al.}{1995}]{young95}
{Young} J.~S.,  {Xie} S.,  {Tacconi} L.,  {Knezek} P.,  {Viscuso} P.,
  {Tacconi-Garman} L.,  {Scoville} N.,    {Schneider} S.,  1995, ApJS, 98, 219

\end{thebibliography}

\newpage
\begin{appendices}
	\begin{center}
		\textbf{\appendixpage}
	\end{center}
\appendix
\section{\textbf{MULTIPLE POINTING RESULTS}}
 \label{amp}

Below we briefly comment on the molecular distribution in each spiral for which we have multiple pointings. \textcolor{black}{In the case of \cg120 we also show an example of the gauss fit to the \hoz\ pointings along the major axis and briefly discuss the fit. } We also discuss the distribution of  H$ \alpha $ which is normally  correlated with CO \citep{komugi07}.  Features of the molecular and  H$ \alpha $ distributions (from GOLDMine) are discussed below:

\begin{figure*}
\begin{center}
\includegraphics[ angle=0,scale=0.60] {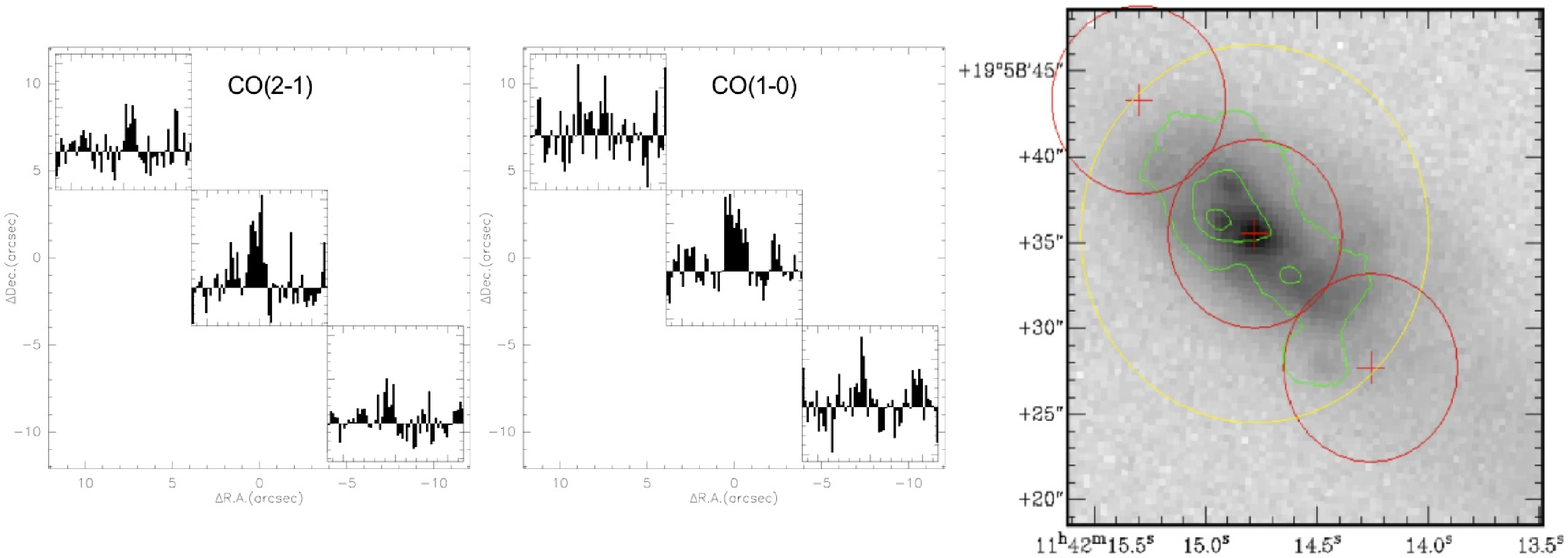}
\vspace{1cm}
\caption{\cg062 \hto\ spectra (left) and \hoz\ spectra (centre), with the (0,0) position  at the galaxy's optical centre;  the $\alpha$ and $\delta$ offsets are in arcseconds. \textcolor{black}{The velocity axes of the individual spectra cover a velocity range of 1300 \km, centred on the optical velocity in Table \ref{pramss},  ($\delta V_{CO}$ = 20 \km) and the $T_\mathrm{mb}$  axes span -5.7 mK to 8.5  mK  and -8.7 mK  to 22.1  mK  for \hoz\ and \hto\ respectively.} The image is an SDSS \textit{r}--band. The yellow circle indicates the size of the 2.6 mm beam at the central pointing position. Red crosses indicate the position of a 1.3 mm observation with a red circle  added to indicate the size of the 1.3 mm beam if  \hto\ was detected at that position. Green contours trace \halpha\ emission from GOLDMine.}
\label{offs62}
\end{center}
\end{figure*}

\begin{figure*}
\begin{center}
\includegraphics[ angle=0,scale=0.49] {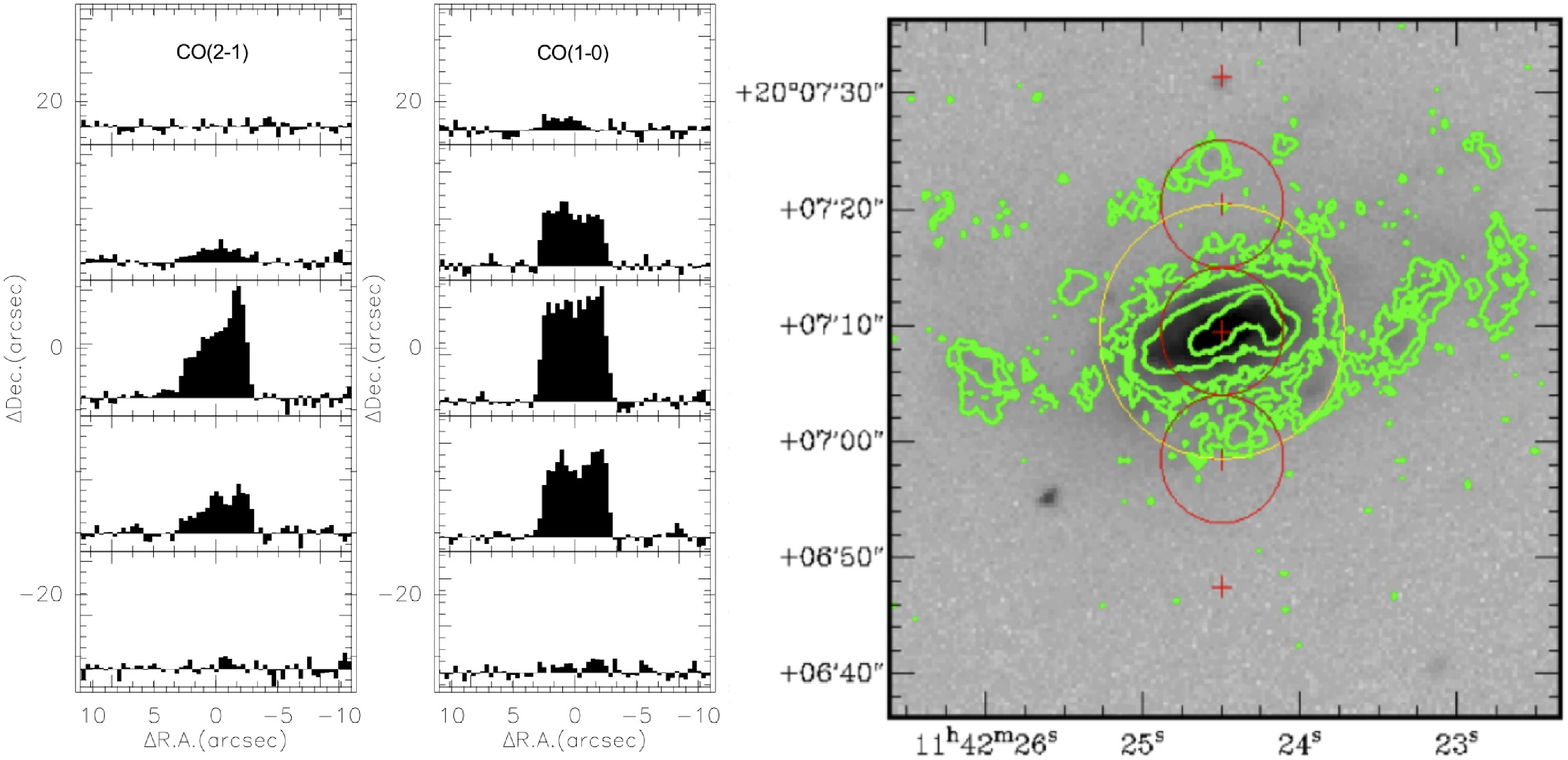}
\vspace{1cm}
\caption{\cg068 \hto\ spectra (left) and \hoz\ spectra (centre), with the (0,0) position  at the galaxy's optical centre;  the $\alpha$ and $\delta$ offsets are in arcseconds.  \textcolor{black}{The velocity axes of the individual spectra cover a velocity range of 1300 \km, centred on the optical velocity in Table \ref{pramss},  ($\delta V_{CO}$ = 20 \km) and the $T_\mathrm{mb}$  axes span -6.0 mK to 51.6  mK  and -16.7 mK to 109.1  mK  for \hoz\ and \hto\ respectively.}  The image is an SDSS \textit{r}--band. The yellow circle indicates the size of the 2.6 mm beam at the central pointing position. Red crosses indicate the position of a 1.3 mm observation with a red circle  added to indicate the size of the 1.3 mm beam if  \hto\ was detected at that position. Green contours trace \halpha\ emission from GOLDMine.}
\label{offs68}
\end{center}
\end{figure*}

\begin{figure*}
\begin{center}
\includegraphics[ angle=0,scale=0.49] {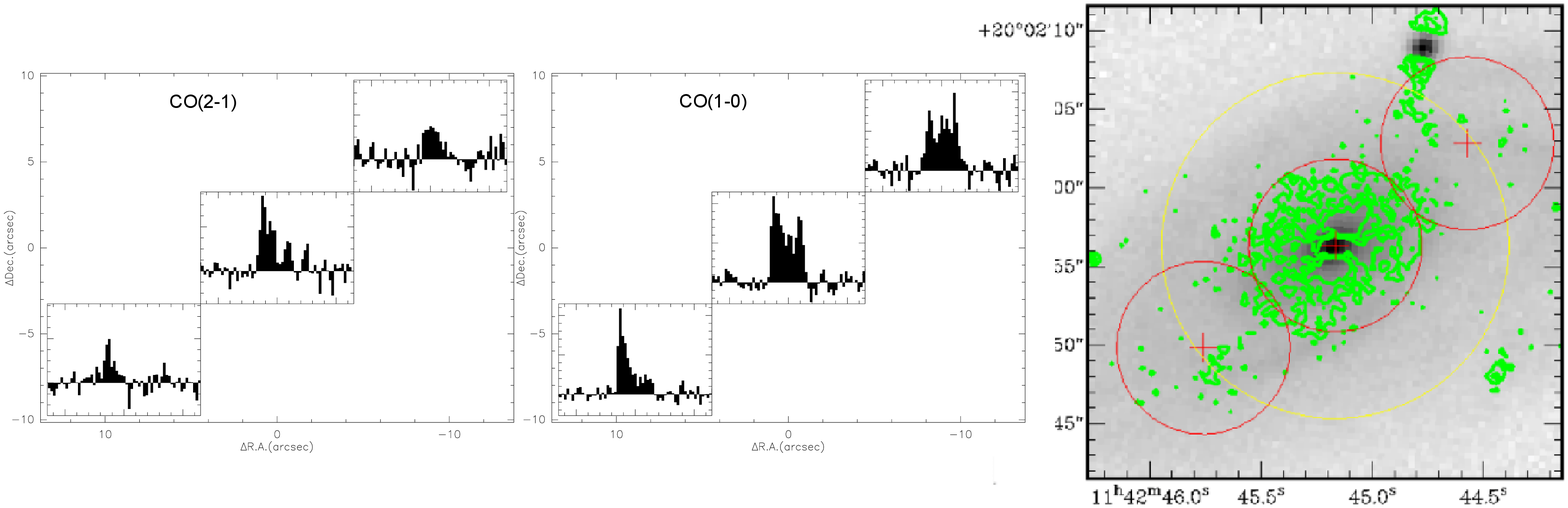}
\vspace{1cm}
\caption{\cg072 \hto\ spectra (left) and \hoz\ spectra (centre), with the (0,0) position  at the galaxy's optical centre;  the $\alpha$ and $\delta$ offsets are in arcseconds.  \textcolor{black}{The velocity axes of the individual spectra cover a velocity range of 1300\km, centred on the optical velocity in Table \ref{pramss},  ($\delta V_{CO}$ = 20 \km) and the $T_\mathrm{mb}$  axes span -5.4 mK to 23.0  mK  and -14.8 mK to 35.8  mK  for \hoz\ and \hto\ respectively.}  The image is an SDSS \textit{r}--band. The yellow circle indicates the size of the 2.6 mm beam at the central pointing position. Red crosses indicate the position of a 1.3 mm observation with a red circle  added to indicate the size of the 1.3 mm beam if  \hto\ was detected at that position. Green contours trace \halpha\ emission from GOLDMine.}
\label{offs72}
\end{center}
\end{figure*}

\begin{figure*}
\begin{center}
\includegraphics[ angle=0,scale=0.65] {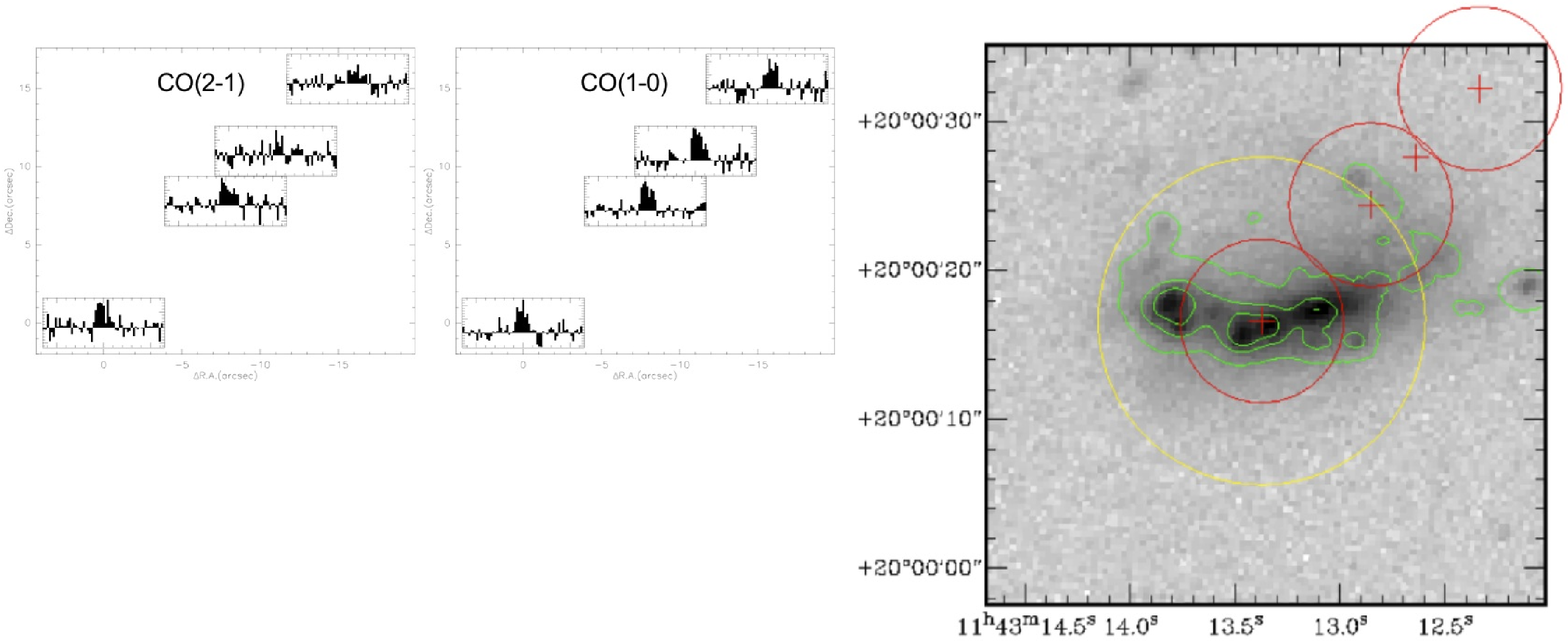}

\vspace{1cm}
\caption{\cg079  \hto\ spectra (left) and \hoz\ spectra (centre), with the (0,0) position  at the galaxy's optical centre;  the $\alpha$ and $\delta$ offsets are in arcseconds.  \textcolor{black}{The velocity axes of the individual spectra cover a velocity range of 1300 \km, centred on the optical velocity in Table \ref{pramss},  ($\delta V_{CO}$ = 20 \km) and the $T_\mathrm{mb}$  axes span -5.7 mK to 12.8  mK  and -13.2 mK  to 19.1  mK  for \hoz\ and \hto\ respectively.}  The image is an SDSS \textit{r}--band. The yellow circle indicates the size of the 2.6 mm beam at the central pointing position. Red crosses indicate the position of a 1.3 mm observation with a red circle  added to indicate the size of the 1.3 mm beam if  \hto\ was detected at that position. Green contours trace \halpha\ emission from GOLDMine.}
\label{offs79}
\end{center}
\end{figure*}

\begin{figure*}
\begin{center}
\includegraphics[ angle=0,scale=0.74] {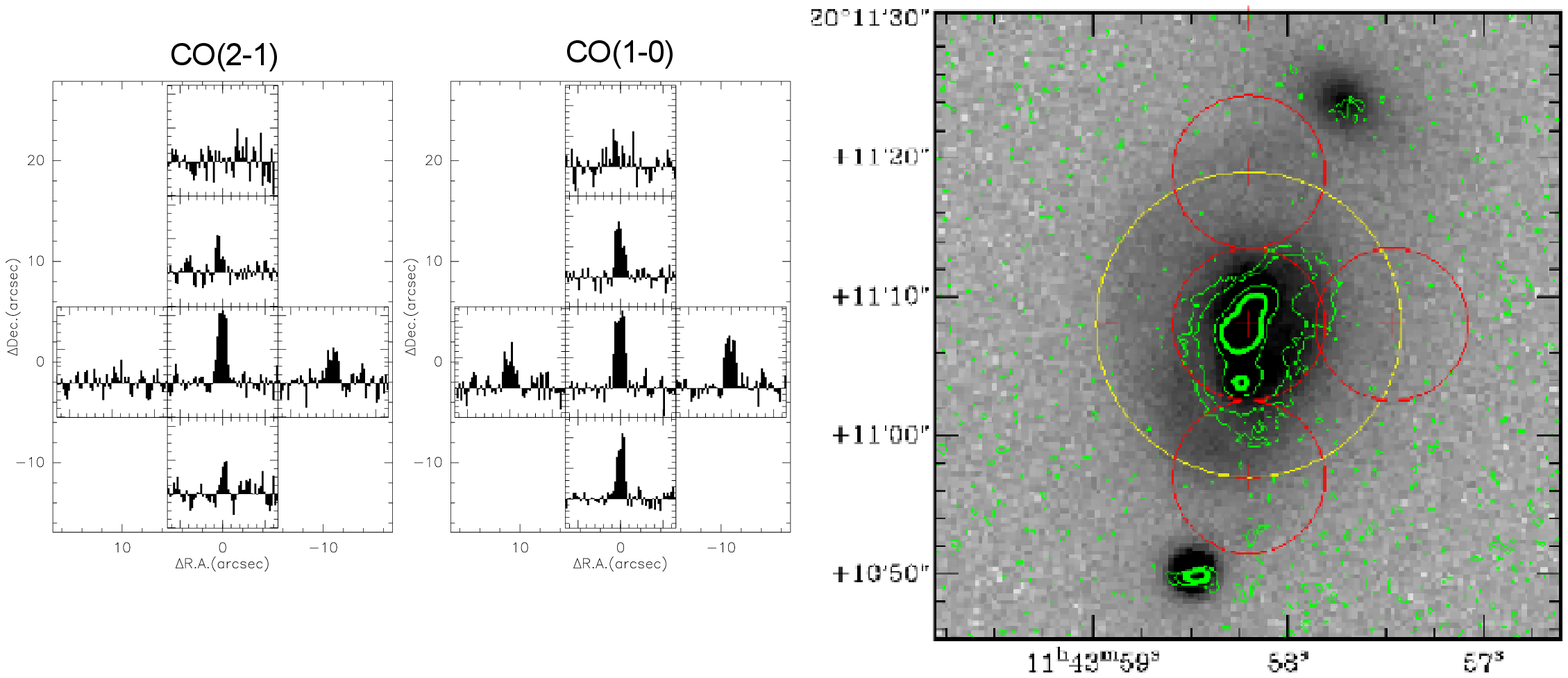}
\vspace{1cm}
\caption{\cg092(S)  \hto\ spectra (left) and \hoz\ spectra (centre), with the (0,0) position  at the galaxy's optical centre;  the $\alpha$ and $\delta$ offsets are in arcseconds.  \textcolor{black}{The velocity axes of the individual spectra cover a velocity range of 1300 \km, centred on the optical velocity in Table \ref{pramss},  ($\delta V_{CO}$ = 20 \km) and the $T_\mathrm{mb}$  axes span -9.6 mK to 26.6  mK  and -20.3 mK to 45.2  mK  for \hoz\ and \hto\ respectively.}  The image is an SDSS \textit{r}--band. The yellow circle indicates the size of the 2.6 mm beam at the central pointing position. Red crosses indicate the position of a 1.3 mm observation with a red circle  added to indicate the size of the 1.3 mm beam if  \hto\ was detected at that position. Green contours trace \halpha\ emission from GOLDMine.}
\label{offs92}
\end{center}
\end{figure*}

\begin{figure*}
\begin{center}
\includegraphics[ angle=0,scale=0.6] {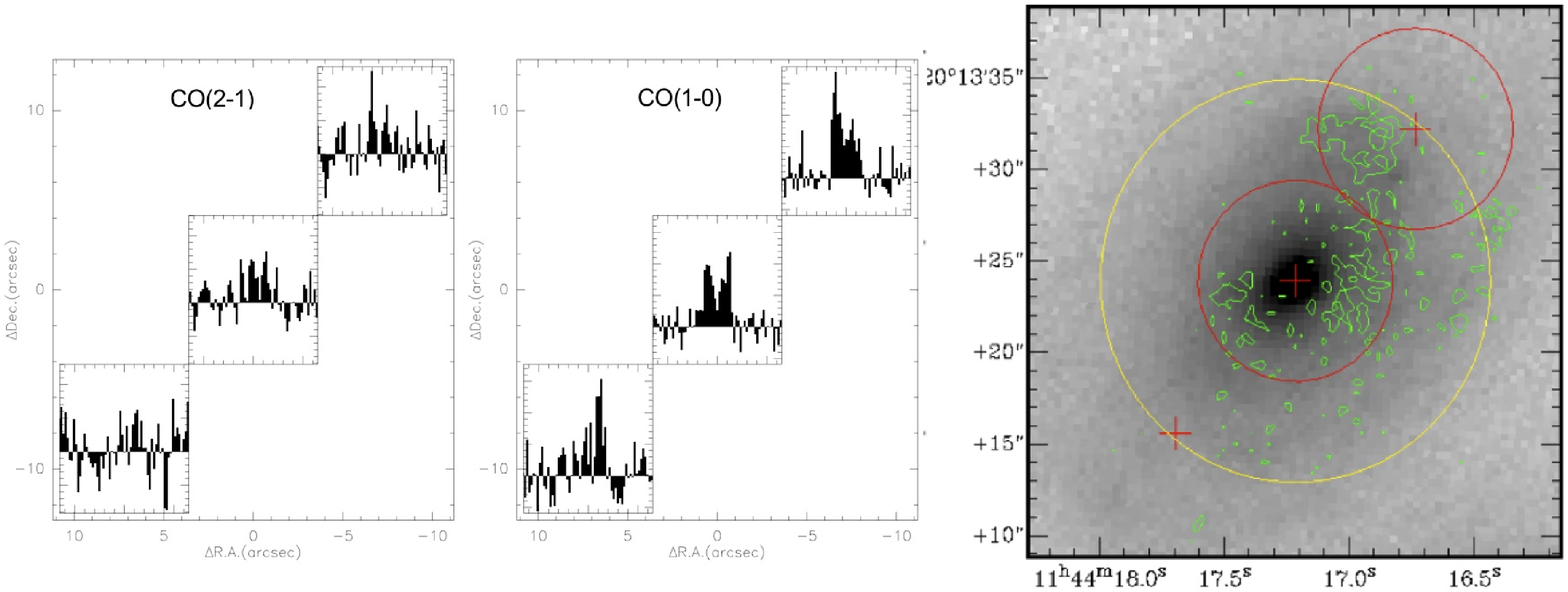}
\vspace{1cm}
\caption{\cg102(N) \hto\ spectra (left) and \hoz\ spectra (centre), with the (0,0) position  at the galaxy's optical centre;  the $\alpha$ and $\delta$ offsets are in arcseconds.  \textcolor{black}{The velocity axes of the individual spectra cover a velocity range of 1300 \km, centred on the optical velocity in Table \ref{pramss},  ($\delta V_{CO}$ = 20 \km) and the $T_\mathrm{mb}$  axes span -5.9 mK to 17.7  mK  and -18.3 mK to 26.0  mK  for \hoz\ and \hto\ respectively.}  The image is an SDSS \textit{r}--band. The yellow circle indicates the size of the 2.6 mm beam at the central pointing position. Red crosses indicate the position of a 1.3 mm observation with a red circle  added to indicate the size of the 1.3 mm beam if  \hto\ was detected at that position. Green contours trace \halpha\ emission from GOLDMine.}
\label{offs102}
\end{center}
\end{figure*}

\begin{figure*}
\begin{center}
\includegraphics[ angle=0,scale=0.54] {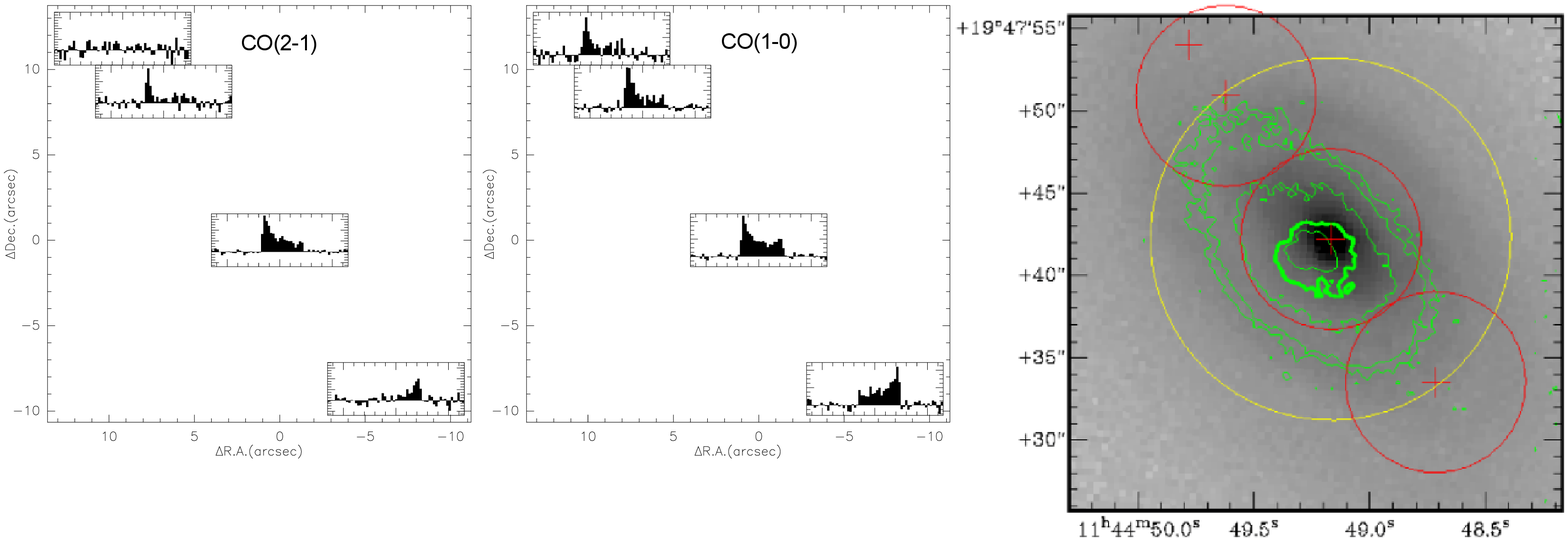}
\vspace{1cm}
\caption{\cg120 \hto\ spectra (left) and \hoz\ spectra (centre), with the (0,0) position  at the galaxy's optical centre;  the $\alpha$ and $\delta$ offsets are in arcseconds.  \textcolor{black}{The velocity axes of the individual spectra cover a velocity range of 1300 \km, centred on the optical velocity in Table \ref{pramss}, ($\delta V_{CO}$ = 20 \km) and the $T_\mathrm{mb}$  axes span -11.6 mK to 48.8  mK  and -27.5 mK to 70.2  mK  for \hoz\ and \hto\ respectively.}  The image is an SDSS \textit{r}--band. The yellow circle indicates the size of the 2.6 mm beam at the central pointing position. Red crosses indicate the position of a 1.3 mm observation with a red circle  added to indicate the size of the 1.3 mm beam if  \hto\ was detected at that position. Green contours trace \halpha\ emission from GOLDMine.}
\label{offs120}
\end{center}
\end{figure*}

\begin{figure*}
\begin{center}
\includegraphics[ angle=0,scale=0.57] {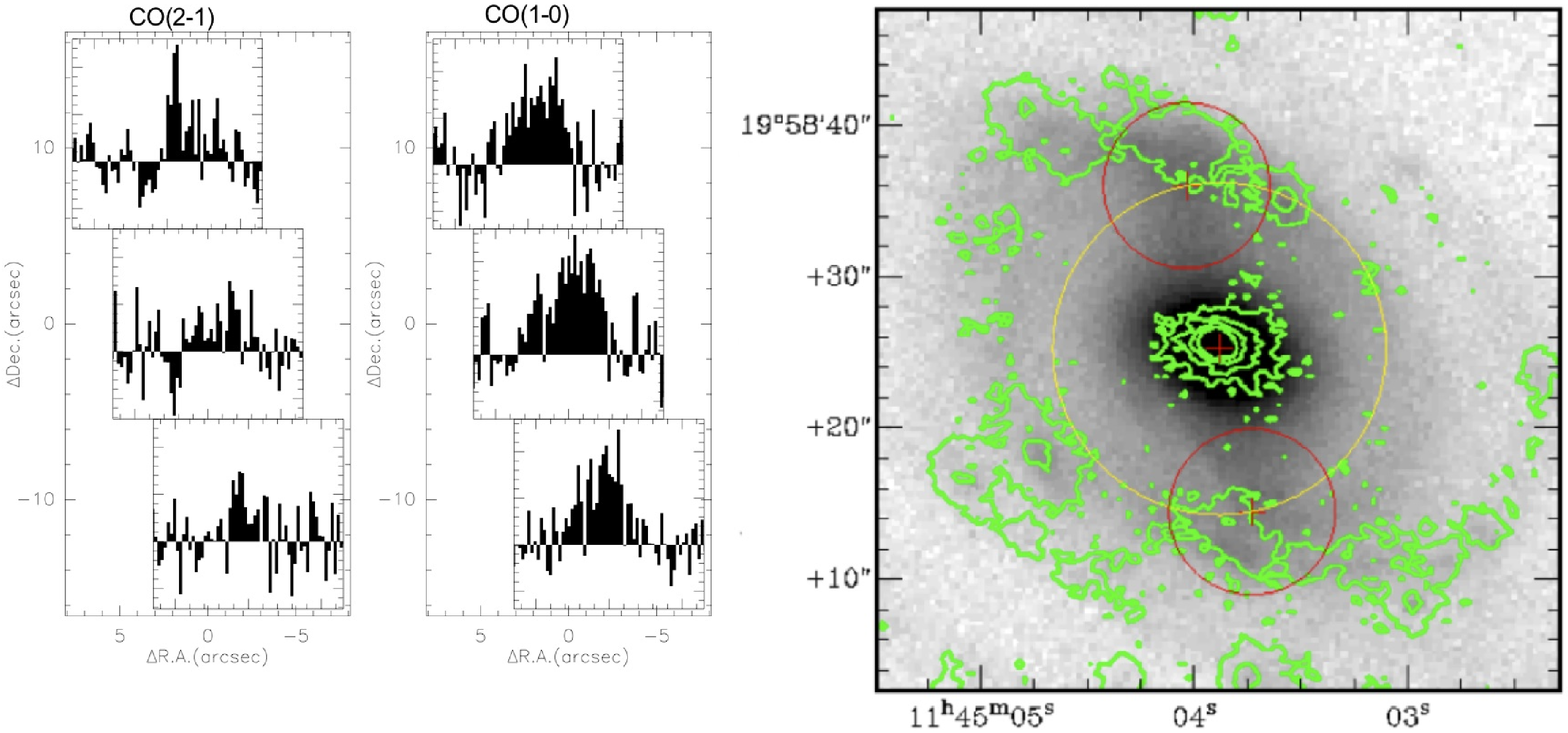}
\vspace{1cm}
\caption{\cg129(W) \hto\ spectra (left) and \hoz\ spectra (centre), with the (0,0) position  at the galaxy's optical centre;  the $\alpha$ and $\delta$ offsets are in arcseconds.  \textcolor{black}{The velocity axes of the individual spectra cover a velocity range of 1300 \km, centred on the optical velocity in Table \ref{pramss},  ($\delta V_{CO}$ = 20 \km) and the $T_\mathrm{mb}$  axes span -6.9 mK to 13.5  mK  and -14.3 mK to 26.0  mK  for \hoz\ and \hto\ respectively.}  The image is an SDSS \textit{r}--band. The yellow circle indicates the size of the 2.6 mm beam at the central pointing position. Red crosses indicate the position of a 1.3 mm observation with a red circle  added to indicate the size of the 1.3 mm beam if  \hto\ was detected at that position. Green contours trace \halpha\ emission from GOLDMine.}
\label{offs129}
\end{center}
\end{figure*}

\begin{figure}
\begin{center}
\includegraphics[ angle=0,scale=0.45] {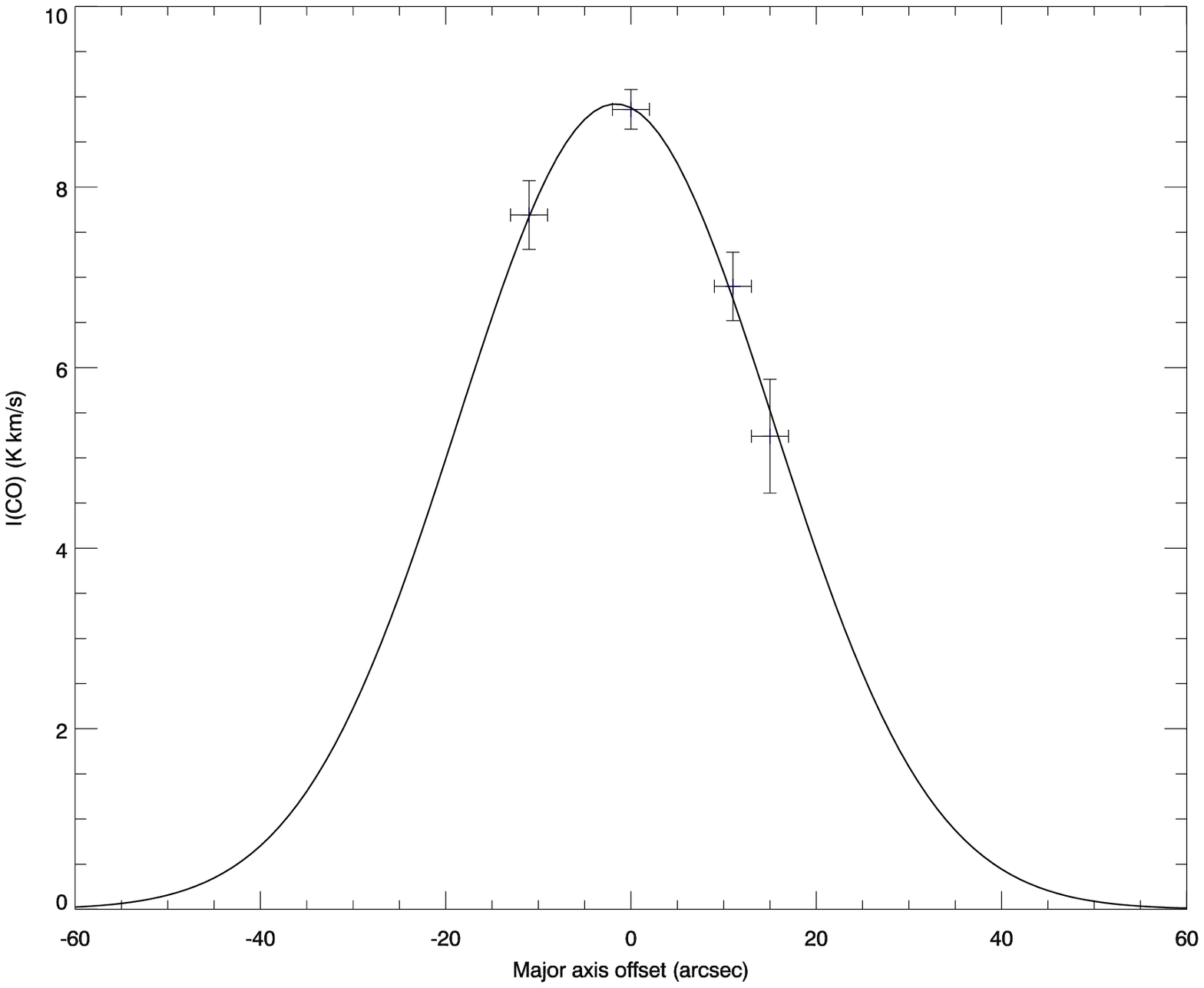}
\vspace{1cm}
\caption{\textcolor{black}{Gauss fit to \hoz\ pointings along the major axis of \cg120. Positional uncertainties are 2 arcsec with  the \hoz\ intensities and their uncertainties  from Table \ref{cosum}.} }
\label{120fwhm}
\end{center}
\end{figure}

\begin{itemize}
\item 
\textbf{\cg062} Optically this peculiar galaxy is strongly asymmetric. The overlay of  \halpha\ contours in Figure \ref{offs62} reveals the \halpha\  to be distributed asymmetrically as well,  strongly favouring the NE. However within the errors there was no indication that the molecular gas distribution was displaced.  

\item
\textbf{\cg068 } 
The \hoz\ and \hto\ spectra in Figure \ref{offs68} suggest a mild asymmetry in the integrated intensity along the minor axis. According to the Gaussian fit, the asymmetry is compatible with an offset of $\sim$ 2 arcsec between the CO peak intensity and the galaxy center. This is of the order of the expected pointing errors.

\item 
\textbf{\cg072} At the central pointing position of this highly inclined ($\sim$ 63\degree) spiral an asymmetric double peak is seen in both the \hoz\ and \hto\ spectra. The \hoz\ spectra at all three positions is detected over a similar range of velocities, although in the SE the distribution of intensity is skewed toward lower velocities compared to the other two spectra (Figure A3). The Gauss fit to the \hoz\ spectra shows the intensity maximum offset by $\sim$3.4 arcsec (1.4 kpc) to the NW of the optical centre, in excess of the pointing uncertainty. But the offset is not confirmed by the \hto\ fit which shows an offset of only 1.3 arcsecs.

Figure \ref{offs72}  shows \halpha\  emission concentrated in the inner $\sim$ 5 arcsec (2 kpc)  of the optical centre. The significantly greater \defhi\ (0.55; Paper I) than \hdos\ deficiency (-0.20) is suggestive of  a prior history involving ram pressure stripping.

\item 
\textbf{\cg079} is optically a highly irregular spiral with indications from its \textit{H}--band (GOLDMine) and 2MASS images that its old stellar population is also perturbed. The galaxy was observed with pointings at 0, 11, 16 and 22 arc- sec along a PA of 315\degree\ from the optical center (Figure \ref{offs79}), following up on earlier observations by \cite{bosel94}. Their \hoz\ spectra suggested, but at low significance, that the CO distribution was offset to the NW of the optical centre. The offset derived from a Gauss fit, including 2 data points from \cite{bosel94}, indicates the intensity maximum lies 15.6 $\pm$ 8.5 arcsec (6 kpc) NW of the optical centre. Additionally the FWHM of the Gaussian fit implies its \hoz\ is distributed more broadly than other spirals in the sample (Figure \ref{gauss}). The detection of \hto\ 22 arcsec (9 kpc) NW of the optical cen- ter (beyond the optical galaxy) provides strong confirmation of this. VLA C--array imaging (angular resolution 15 arcsec) by  \cite{hota07} and our VLA D--array observations (Paper I) both show the \hi\ intensity maximum offset at a similar position to the \hoz\ intensity maximum derived from the gauss fit.

The EW(\halpha +[NII]) of 130\AA\ is one of the largest in the Coma and A 1367 sample \citep{ipara02} which is consistent with enhanced SF during an interaction. Also the \halpha\ contours in Figure \ref{offs79}, if not the result of dust extinction, imply the bulk of the current star formation is occurring co-spatially with the optical disk $\sim$5--10 kpc SE of the \hi\ and molecular gas intensity maxima. At the current SFR (1.06 \msolar\ yr$^{-1}$) the spiral's \hdos\ will be depleted in $\sim$ 9.6 x 10$^8$ yr. UV from GALEX \citep{cort05} indicates strong star formation in the recent past East of the current (\halpha) peak. These star formation signatures point to a recent triggering of the starburst. A possible explanation for the asymmetric gas distributions and starburst is a tidal encounter with a close neighbour. A NED search within 5 arcmin ($\sim$ 125 kpc) in the velocity range 6585 \km\ to 7585 \km\ of \cg079 reveals four	galaxies	which	and could potentially have interacted with it within a time scale of 10$^8$ yr. But in all four cases, including \cg073, the perturbation parameter\footnote{ $p_{gg}= \frac{(M_{comp} / M_{gal}) }{(d/r_{gal})^3} $ where $M_{gal}$ and $M_{comp}$ are the masses of the galaxy  and companion respectively, d is the separation and r is the galaxy disk radius \citep{byrd90}, where values of $p_{gg}$ $>$ 0.1 likely lead to tidally induced starformation. }  $p_{gg}$ is at least an order of magnitude below that expected to produce detectable molecular gas perturbations in \cg079; none of the four shows a strongly disturbed optical morphology.

A prominent feature of this galaxy is its $\sim$ 75 kpc \halpha\ and radio continuum tails oriented away from the cluster centre which is taken as evidence that this spiral is suffering significant ram pressure stripping \citep{gava01b,gava87}. It remains unclear whether a strong tidal interaction is required to account for the starburst. The displacement of the molecular gas and the tails could in principle in part be due to ram pressure removing the gas more efficiently once it is loosened by a tidal interaction.

\item 
\textbf{\cg092(S)} This spiral has a highly disturbed optical morphology, including an $\sim$8 arcsec (3 kpc) N--S bar--like nucleus, also seen in \halpha\ (EW(\halpha) = 28  \AA, Figure \ref{offs92}). The velocity separation of 1139 \km\ between \cg092(S) and \cg092(N), projected 16 arcsec (6.5 kpc) to the NW, (p$_{gg}$ of 0.08) together with the optical bridge linking them with a tidal interaction which may have induced the central bar in \cg092(S). Although in general the shorter duration of tidal interactions in clusters is expected to mitigate against bar formation \citep{bosel06a}.

Inspection of the \hoz\ spectra in Figure \ref{offs92} suggests an E--W asymmetry in the distribution of CO gas perpendicular to the optical bar. However, the \hoz\ Gauss fits (Table \ref{gauss}) indicate the intensity maximum is offset to the west of the optical centre (1.9 arcsec) which is the order of the pointing uncertainty.


 
\item 
\textbf{\cg102(N)}  The  \hoz\   spectra (Figure \ref{offs102})  show an asymmetry in the distribution of CO gas along the major axis. The   \hoz\ Gauss fit (Table \ref{gauss}) also indicates  the intensity maximum is offset 11.1 $\pm$11.2 arcsec (4.5 kpc)  to the NW  of the optical centre, but the uncertainty is of the same order as the offset. There is strong evidence this galaxy is tidally interacting with its close companion, \cg102(S), projected 0.42 arcmin (10 kpc)  to the SW. The assertion is based on: (i) the tidal bridge between two galaxies seen in the 2MASS $J, H$ and $K$ images of the pair, (ii) p$_{gg}$ = 0.3  and (iii) the velocity separation of 4\,\km. The EW(\halpha) for \cg102(N) is low at $\sim$ 2 \AA\ but the weak \halpha\ emission  distributed towards the north and west  may be indicative of  a more dispersed molecular distribution and is consistent with the large FWHM of the  \hoz\ Gauss fit of 56.9$\pm$5.2 arcsec (Figure \ref{gffwhm}).

\item 
\textbf{\cg120} As Table \ref{calc} shows, despite having a \dhi\ $ \geq $0.78 this spiral  is  normal in molecular content (\hd\ = -0.2).  The  mean position of the intensity maximum of  both the \hoz\ and \hto\ Gaussian fits to detections along the major axis is consistent with the optical centre to within  the pointing uncertainties, implying an approximately \textcolor{black}{symmetric molecular gas distribution. This is illustrated in Figure  \ref{120fwhm} which shows the Gaussian fit to four \hoz\ pointings along the major axis of the galaxy. The offset from the optical is  -1.7$\pm$ 1.0 arcsec which is less than the pointing uncertainty of $\sim$ 2 arcsec. The FWHM of the fit is 40.0$\pm$1.8 arcsec which is used to assess the size of the molecular disk in section \ref{maps}. The weak}  \halpha\ emission (EW(\halpha) = 4 \AA) also has a symmetrical distribution  concentrated near the optical centre (Figure \ref{offs120}). This Sa spiral being optically undisturbed, CO rich and highly \hi\  deficient displays the characteristics expected in a spiral which has experienced significant ram pressure stripping \citep{lucero05}.


\item 
\textbf{\cg129(W)} is a large  optically undisturbed \hi\ rich spiral. The spectra in Figure \ref{offs129}, suggest a symmetric CO distribution, which is  confirmed by the \hoz\ Gauss fit which had an the offset from the optical centre smaller than the pointing uncertainty. The \halpha\ distribution (Figure \ref{offs129}) is  concentrated near the optical centre although there is extensive \halpha\ emission  in the outer  spiral arms.  The difference between the \hdos\ mass determined from the IRAM and Kitt Peak \hoz\ implies $\sim$ 50\% of the \hdos\  is located in these spiral arms. Overall the  \hdos\ deficiency (0.09 from Kitt Peak flux) and the  \hi\ deficiency (-0.07) indicates this large galaxy has normal \hi\ and \hdos\ content.  

 \end{itemize}

 \vspace{2.5 cm}
 
\section{\textbf{CO INTENSITIES, FLUXES AND \hdos\ MASS}}
\label{cofm}
In this appendix we summarise the basis of our calculation of \hdos\ mass from integrated intensities using  \ico\ on the $T_\mathrm{mb}$ scale.
 
 \subsection{preliminaries}
  $T_\mathrm{mb}$ is related to $T_A^*$ as follows: 
\begin{equation}
\hspace{7 mm} T_\mathrm{mb}  = \frac{T_A^*}{ \eta_{mb}}
\end{equation}
\begin{equation}
\hspace{7 mm} T_R^*  = \frac{T_A^*}{ \eta_\mathit{fs}}
\end{equation}
Where \\
$\eta_{mb}$ = the beam efficiency;\\
$\eta_\mathit{fs}$ = forward scatter and spillover.\\
In the equations below we allow for the modification of the standard conversion factor ($X_{CO}$) between integrated intensities and the number of \hdos\ molecules N(\hdos) by defining a correction factor, $ \chi_{co}$, to the standard conversion:  
 
 \begin{equation}
\hspace{7 mm}\chi_{co} = \frac{X_{CO}^H}{2.3 \times 10^{20}}
\end{equation}
Where: \\
 $ \chi_{co}$ = the correction factor applied to the standard  conversion factor $X_{CO}$  \citep[taken as $X_{CO} =  2.3 \times $10$^{20}$ mol cm$^{-2}$ (K\,\km)$^{-1}$;][]{strong88,Polk88} to bring the conversion factor up to the \textit{H}--band luminosity dependent $X$--factor from \cite{bosel02} or  $ \chi_{co} = 1$ if the standard conversion factor is to be applied;\\
 $X_{CO}^H$ = the $L_H$ dependent conversion factor from \cite{bosel02}. \\
 
The \hdos\ column density is then: 

\begin{equation}
\hspace{7 mm}N(\mathrm{H_2}) = 2.3 \times 10^{20}\, \chi_{co}\,\sum_{i} T_\mathrm{mb}^i \Delta\,V\,[\mathrm{mol\,cm^{-2}}]
\end{equation}
Where: \\
 $T_\mathrm{mb}$ = the Rayleigh--Jeans approximation of the main beam brightness temperature referred  to as the surface brightness;\\
 $\sum_{i} T_\mathrm{mb}^i \Delta\,V $ = the summation of $T_\mathrm{mb}$ over all the observed spectral channels, often denoted as \ico.\\ 
 
\noindent
N($H_2$) expressed in  \msolar\ pc$^{-2}$ becomes:

 \begin{equation}
\hspace{7 mm}N(\mathrm{H_2}) = 3.67\, \chi_{co}\, \sum_{i} T_\mathrm{mb}^i \Delta\,V\,[\mathrm{M_{\odot}\,pc^{-2}}]
\label{eqa_5}
\end{equation}
The total gas (inclusive of a correction for a 1.36 He mass fraction) then becomes\\
\begin{equation}
\hspace{7 mm}N(\mathrm{H_2 + He}) = 5.0\, \chi_{co}\,\sum_{i} T_\mathrm{mb}^i \Delta\,V\,[\mathrm{M_{\odot}\,pc^{-2}}]
\label{eqa4}
\end{equation}

 \subsection{Flux density and telescope gain}
Flux density is the quantity measured by a telescope and is beam (and hence wavelength)  dependent
\begin{equation}
\hspace{7 mm}S_i = \frac{2\, k\, T^i_\mathrm{mb} }{ \lambda^2}  \iint_{\mathrm{mb}}  P  \, d\Omega
\end{equation}
 
Where\\
$S_i$ = flux density [Jy beam$^{-1}$] in velocity channel $i$ of width \textcolor{black}{$\delta V_{CO}$ }\\
$\lambda$ = wavelength [meters], and $k$ is the Boltzmann constant. \\

Here it is assumed that the emission is more extended than the beam, but can be thought of as constant across the beam. If the beam can be well approximated by a 2--D Gaussian, normalised at its peak to 1.0, and the FWHM is $(B_\alpha\,,B_\delta)$  in arcsec and $(\Delta\,l\,,\Delta\,m)$ in radians, this can be written as:

\begin{equation}
\hspace{7 mm}S_i = \frac{2\, k\, T^i_\mathrm{mb} }{ \lambda^2} \,1.13\,\Delta\,l\,\Delta\,m
\end{equation}

In the specific case of \hoz\ observed with the IRAM 30--m telescope the FWHM of the  beam $B_\alpha = B_\delta$ = 22\prin\  and
\begin{equation}
\hspace{7 mm}S \mathrm{[Jy\, beam^{-1}]} = 1.085 \times\ 10^{-2}\, T_\mathrm{mb}  \mathrm{[K]} \,B_\alpha \, B_\delta
\label{equ2}
\end{equation}

or a telescope gain  of $G = 4.95$\,Jy\,K$^{-1}$.

\subsection{Mass of resolved sources }\
For sources that are more extended than the beam and for which a map is available for the whole source so that a summation can be carried over its full surface area, it follows from equation \eqref{eqa_5} that :

\begin{equation}
\hspace{7 mm}M(\mathrm{H_2}) = 3.67\, \chi_{co}\, \sum_{x,y}\,\sum_{i} T_\mathrm{mb}^i \Delta\,V\,\Delta\,x\,\Delta\,y\,[\mathrm{M_{\odot}}]
\label{eqa_15}
\end{equation}
Where $\Delta x$ and $\Delta y$ are the dimensions of ``rectangular" pixels, measured in pc. For a distance to the source D\,[Mpc] and rectangular pixels of angular size (in arcsec) of $(\Delta \alpha, \Delta \delta)$, this becomes:

\begin{equation}
\hspace{7 mm}M(\mathrm{H_2}) = 86.3\, \chi_{co}\, D^2 \,  \sum_{\alpha,\delta} T_\mathrm{mb}^i \, \Delta\,V \,\Delta\,\alpha\, \Delta\,\delta\,[\mathrm{M_{\odot}}]
\label{eqa_18a}
\end{equation}
and
\begin{equation}
\hspace{7 mm}M(\mathrm{H_2 + He}) = 117\, \chi_{co}\, D^2 \,  \sum_{\alpha,\delta} T_\mathrm{mb}^i \, \Delta\,V \,\Delta\,\alpha\, \Delta\,\delta\,[\mathrm{M_{\odot}}]
\label{eqa_18b}
\end{equation}

\subsection{Unresolved sources}
\label{cofm4}
For unresolved sources, the spatial distribution of the intensity as measured by the telescope resembles the beam shape. On the assumption of a point like source and a Gaussian beam, the mass of the source can be calculated from the central pointing as:

\begin{equation}
\hspace{7 mm}M(\mathrm{H_2}) =7950\, \chi_{co}\, D^2 \,  \sum_{i} S_i \, \Delta\,V \,[\mathrm{M_{\odot}}]
\label{eqa_20}
\end{equation}
\begin{equation}
\hspace{7 mm}M(\mathrm{H_2 + He}) =1.06 \times\ 10^4 \, \chi_{co}\, D^2 \,  \sum_{i} S_i \, \Delta\,V \,[\mathrm{M_{\odot}}]
\label{eqa_21}
\end{equation}
Where
\begin{equation}
\hspace{7 mm} \sum_{i} S_i \, \Delta\,V = G [\mathrm{Jy\,K^{-1}] \times I(CO) [K km\,s^{-1}}]
\label{eqa_22}
\end{equation}

Here $S_i$ is the flux density (Jy/beam) in the ith channel and G is the telescope gain. For the IRAM 30--m telescope at the time of our observations, G = 4.95 Jy/K.

The latter relations are independent of the beam size as long as the source is unresolved.
 
\section{\textbf{COMPARISON WITH KITT PEAK 12--m CO OBSERVATIONS}}
\label{appc}

Ten of the spirals in our sample have previously been observed in \hoz\ with the Kitt Peak 12--m telescope \citep{bosel97}. In this appendix we will compare our IRAM 30--m \hoz\ observations with those results. If indeed, as we argue in the main text, the sources are unresolved in the Kitt Peak 55 arcsec and the IRAM 30--m 22 arcsec beams, it follows from Appendix \ref{cofm4} that the comparison that involves the least assumptions is that between the fluxes, S [Jy\,km\,s$^{-1}$]. If unresolved, it suffices to deal with the central pointing (for those targets for which multiple pointings were obtained).

We proceeded as follows. We converted the Kitt Peak intensity \ico\ from the $T_R^*$ to $T_\mathrm{mb}$ via $T_\mathrm{mb}$ = $T_R^*/0.82$.  We then converted  $T_\mathrm{mb}$ from both telescopes  to  fluxes S [Jy\,km\,s$^{-1}$] by multiplying with the Gain of the respective telescopes. The results are listed in Table~\ref{fluxes}.

The comparison with previous results is shown in Figure \ref{fluxx}. The Kitt Peak values for \cg062, \cg063, and \cg102 are upper limits as are our IRAM values for \cg063 and \cg082. The figure is quite revealing. Although there is reasonable agreement between the fluxes from the two telescopes for each spiral, in the sense that they show the same general trend, the IRAM fluxes are consistently lower by about 50\%. There is a lager discrepancy between the values for \cg072, \cg082, \cg121, and  \cg129.

 \cg082 was a non--detection with IRAM and, as mentioned in the main text, an independent observation will be required to settle the issue which of the two measure- ments is correct. For  \cg072  we consider the IRAM result to be more reliable as the S/N is 23.5 compared to the Kitt Peak S/N of 3.6 and the optical size of the galaxy is similar to other galaxies where the agreement between the flux of the two telescopes was to within a factor of two, with the measurements having similar errors. There is a substantial mismatch between both instruments in the case of \cg129. This is not surprising as the assumption of the target being unresolved is clearly violated. The IRAM central pointing picks up only a fraction of the total flux.


Can we estimate by how much we underestimate the \hdos\ mass? Comparing the \hto\ with the \hoz\  spectra gives some indication as to the extent of the CO emission. The former often show a lack of emission on either the high or low velocity side of the profile, indicating it is definitely sampling a smaller area and that CO at higher velocities lies beyond the 11 arcsec (4.5 kpc) FWHM of the \hto\ beam. The median FWHM of the Gaussian fits to the \hoz\ emission projected across the spirals with multiple pointings (excluding the two probable interacting cases) was 26.4 arc seconds, slightly larger than the 22 arcsec FWHM of the 2.6 mm beam (Table \ref{gauss}), suggesting that for a typical spiral at the distance of Abell 1367 the bulk of the \hoz\ emission is detected in the \hoz\ central pointing. Only in the case of the larger spirals \cg120 (FWHM 41.3 $\pm$ 2.0 arcsec) and \cg129(W) (32.2 $\pm$ 2.1 arcsec) up to 50 \% of the \hoz\ emission could lie beyond the FWHM of the 2.6 mm central pointing.


\cite{fuma09} derive the radial distribution of CO in spirals from a spatially resolved study of CO and \hi\ in a sample of Virgo galaxies. Extrapolating their radial CO density profiles suggests that up to 30 \% of a spiral's CO emission could lie beyond the IRAM 30--m FWHM at 2.6 mm. Furthermore, the HERACLES survey of CO in a sample of nearby spirals \citep{leroy09} found their CO disks had exponential scale lengths  of 2--3\,kpc. So, provided  molecular gas in cluster spirals has a similar exponential radial profile centred on the optical nucleus, the bulk of the CO emission is expected to fall within the IRAM 30--m 2.6 mm   beam. Moreover if the observed  truncation of  star formation in the stellar disks of cluster spirals, as traced in \halpha\    \citep{koop04,bosel06b},  is mirrored in their molecular gas disks, as seems likely, then the fraction of a cluster spiral's  molecular gas within the 2.6\,mm beam will  be higher than for an equivalent galaxy in the field.

\textcolor{black}{Seven of the galaxies that we detect in \hoz\ were observed  with the Kitt Peak telescope (six of them detected) . For these spirals we calculate the Kitt Peak--to--30m ratio of velocity-integrated fluxes (in Jy \km) measured towards the centre, which depends on the extent of the emission. In each case we compare the measured ratio with the ratios expected from the simulated disc (see Section \ref{maps}) assuming different scale lengths. The results of this test are presented in Figure. \ref{kp30}. We assume a typical 0.13 dex uncertainty that combines the typical observational error \textcolor{black}{($\sim$ 25\%)  due to noise in the spectra from which the intensities are measured } and a $\sim$ 20\% uncertainty in the relative calibration of the two telescopes. We find that all ratios are compatible with the scale lengths  found in nearby unperturbed galaxies. The ratio in \cg121 corresponds to a larger than expected disc ($\sim$ 0.6\textcolor{black}{$r_{25}$}), but this evidence is marginal given the large uncertainties.}

In summary, our IRAM 30--m fluxes are likely to be systematically lower by about 30\% and up to 50\% in extreme cases \textcolor{black}{compared to those from Kitt Peak. These differences are well within the 0.4\,dex \hd\  threshold used in the paper. However, clearly} the best way to measure the true CO flux and derive the \hdos\ mass,  while preserving the angular resolution,   is by fully mapping each object with a focal plane array, e.g with the IRAM 30--m -- HERA instrument.

\begin{table} 
\centering
\begin{minipage}{90mm}
\caption{IRAM--30m and Kitt peak 12--m \hoz\ fluxes}
\label{fluxes}
\begin{tabular}[h]{@{}lllll@{}}
\hline
ID& Intensity & Intensity & Flux & Flux \\
& IRAM & KP & IRAM & KP \\ 
& [$T_\mathrm{mb}$] & [$T_\mathrm{mb}$] & [Jy \km] & [Jy \km] \\
\hline\\
97062 & 1.26 & $\leq$0.28 &\,\,\,6.63 &\,\,\,$\leq$9.23 \\
97063 & $\leq$0.43 & $\leq$0.21 &\,\,\,$\leq$2.26 &\,\,\,$\leq$6.82 \\
97072 & 4.16 & 1.26 & 21.90& 41.33 \\
97082 & $\leq$0.34 & 0.55 &\,\,\,$\leq$1.79 & 18.06 \\
97092(S) & 3.37 & 0.77 & 17.74 & 25.28 \\
97093 & 1.35 & 0.35 &\,\,\,7.11& 11.64 \\
97102(N) & 2.35 & $\leq$0.32 & 12.37& $\leq$10.43 \\
97114 & 1.94 & 0.45 &10.21 & 14.85 \\
97121 & 1.58 & 0.74 &\,\,\,8.32 & 24.48 \\
97129(W) & 4.05 & 1.27 & 21.32 & 41.73 \\
\hline
\end{tabular}
\end{minipage}
\end{table}

\begin{figure}
\begin{center}
\includegraphics[ angle=0,scale=0.45] {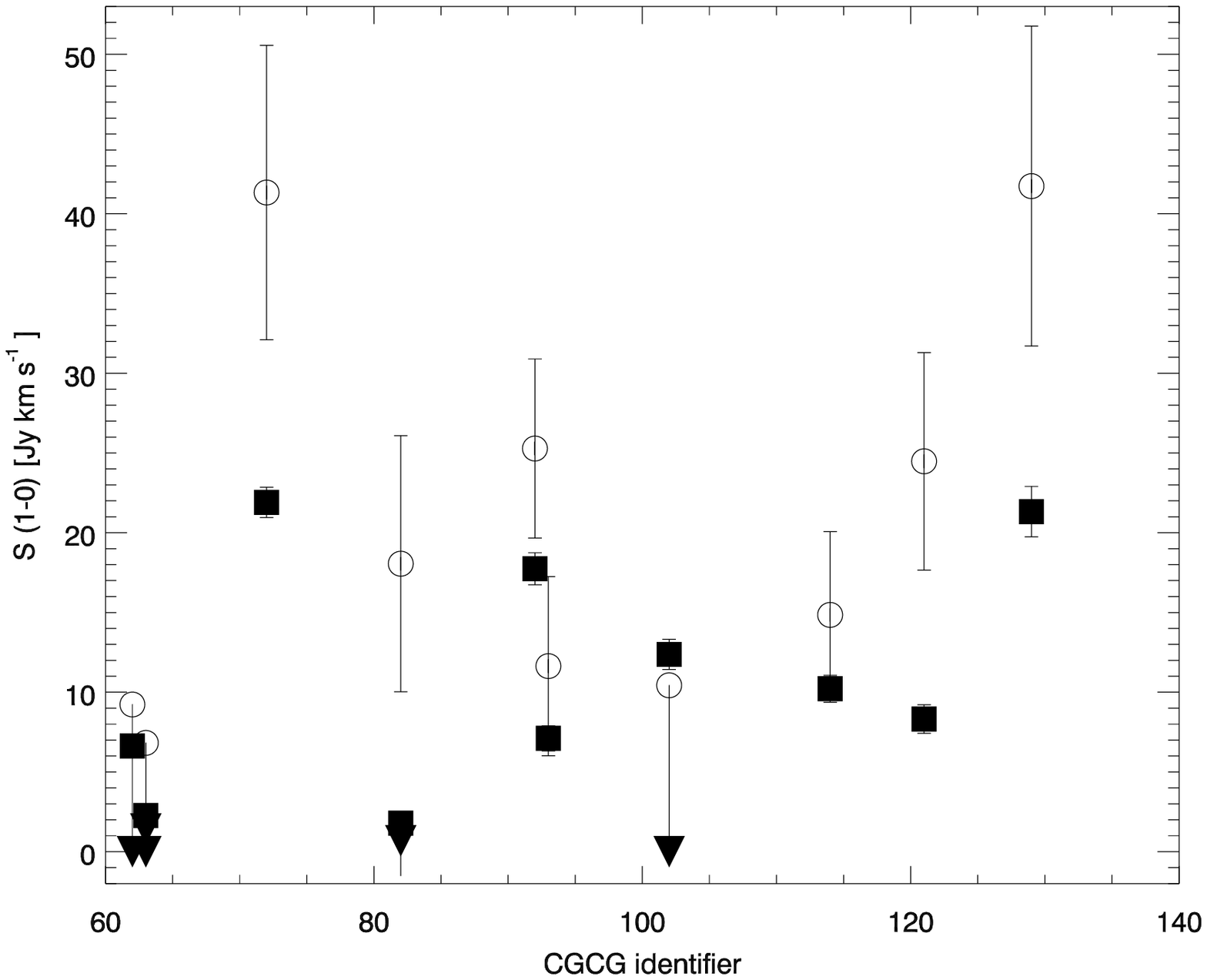}
\vspace{1cm}
\caption{Comparison of \hoz\ fluxes in Jy from IRAM 30--m (filled squares) and \citep{bosel97}  Kitt Peak 12--m (open circles) observations . The x--axis values are the \cg catalogue numbers for the spirals.}
\label{fluxx}
\end{center}
\end{figure}

\begin{figure}
\begin{center}
\includegraphics[ angle=0,scale=0.43] {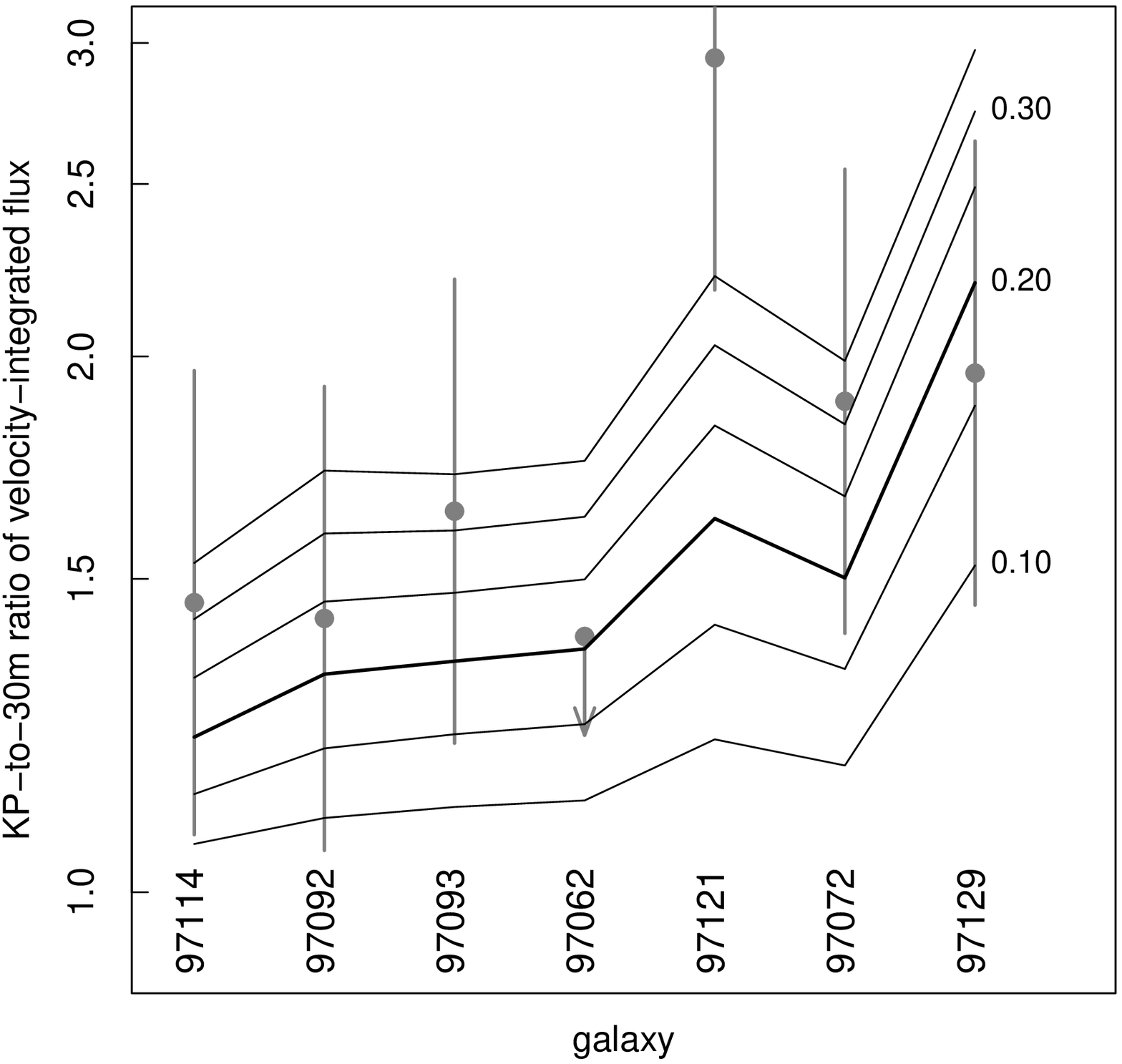}
\vspace{1cm}
\caption{KP--to--30m ratio of velocity-integrated flux of  the galaxies that we detect and were previously observed by Boselli et al. (1997). The galaxies are sorted from left to right in increasing order of optical diameter. The grey dots and error bars indicate measured ratios and their typical $\pm$1$\sigma$ uncertainty (0.13 dex). The solid lines join the values expected from the models assuming CO scale lengths from 0.10 to 0.35 times \textcolor{black}{$r_{25}$} in steps of 0.05\textcolor{black}{$r_{25}$}, as indicated on the right-hand side. }
\label{kp30}
\end{center}
\end{figure}

\end{appendices}

\end{twocolumn}

\end{document}